\documentclass[twocolumn,preprintnumbers,amsmath,amssymb,]{revtex4}

\usepackage{graphicx}% Include figure files
\usepackage{dcolumn}% Align table columns on decimal point
\usepackage{bm}% bold math
\usepackage{xcolor}

\newcommand{\Dcal}{\mathcal{D}}

\newcommand{\mb}[1]{\boldsymbol{#1}}
\newcommand{\br}{\mb{r}}
\newcommand{\bx}{\mb{x}}

\newcommand{\braket}[2]{\langle{#1}|{#2}\rangle}

\newcommand{\brakket}[3]{\langle{#1}|{#2}|{#3}\rangle}

\newcommand{\DM}[2]{| {#1} \rangle\langle{#2} |}

\newlength{\back}

\allowdisplaybreaks[2]

\begin{document}

\title{On derivatives of the energy with respect to total electron number and orbital occupation numbers. A critique of Janak's theorem.\footnote{This paper honors the memory of Dieter Cremer and his penetrating analysis of fundamentals of DFT}}
\author{E. J. Baerends}
\affiliation{Vrije Universiteit, Amsterdam, The Netherlands}

\date{\today}

\begin{abstract}
\noindent
The relation between the derivative of the energy with respect to occupation number and the orbital energy, $\partial E/\partial n_i = \epsilon_i$, was first introduced by Slater for approximate total energy expressions such as Hartree-Fock and exchange-only LDA, and his derivation holds for hybrid functionals as well. We argue that Janak's extension of this relation to (exact) Kohn-Sham density functional theory is not valid. The reason is the nonexistence of systems with noninteger electron number, and therefore of the derivative of the total energy with respect to electron number, $\partial E/\partial N$.  How to handle the lack of a defined derivative $\partial E/\partial N$ at the integer point, is demonstrated using the Lagrange multiplier technique to enforce constraints. The well-known straight-line behavior of the energy as derived from statistical physical considerations \cite{PerdewParrLevyBalduz1982} for the average energy of a molecule in a macroscopic sample (``dilute gas") as a function of average electron number is not a property of a single molecule at $T=0$. One may choose to represent the energy of a molecule in the nonphysical domain of noninteger densities by a straight-line functional, but the arbitrariness of this choice precludes the drawing of physical conclusions from it.  
\end{abstract}

\maketitle

%%%%%%%%%%%%%%%%%%%%%%%%%%%%%%%%%%%%%%%%%%%%%%%%%%%%%
% Section INTRODUCTION:  sec:introduction
\section{Introduction}\label{sec:introduction}
%%%%%%%%%%%%%%%%%%%%%%%%%%%%%%%%%%%%%%%%%%%%%%%%%%%%%%
Fractional occupation numbers have remained popular ever since Slater introduced his expressions of the Hartree-Fock (HF) energy \cite{SlaterMann1969} and of the X$\alpha$ energy (exchange-only LDA, XLDA) using explicit occupation numbers. His relation \cite{Slater1974Vol4} 
\begin{align}\label{eq:SlaterRelation}
&\text{Slater: }\partial E^{model}/\partial n_i=\epsilon^{model}_i  \\
&model \text{ = HF, X$\alpha$, LDA, GGA, hybrid}, \notag
\end{align}
holds for the mentioned model total energies. Slater applied his derivation explicitly to Hartree-Fock and to exchange-only LDA (X$\alpha$), but his method only relies on the application of the chain rule to density and density matrix dependent functionals, so is valid for all cited cases. Slater used relation \eqref{eq:SlaterRelation} to solve a problem that plagued his X$\alpha$ scheme and all (semi-)local density functional approximations (LDFAs) since then: the orbital energy of the highest occupied orbital is not close to minus the ionization energy $-I$. In the case of HF it can be shown that the HOMO orbital energy $\epsilon_H^{HF}$ \textit{is} close to $-I$, acording to Koopmans' frozen orbital approximation for the ion,  $\epsilon_H^{HF}=-I^{HF}(froz. orb.)$. The frozen orbital approximation leads to $\epsilon^{HF}_H$ being a bit lower than the exact $-I$ (often about 1 eV). But for the local and semi-local density functional approximations (LDFAs) like LDA and GGAs it is always some 4 - 6 eV too high. Slater introduced his so-called transition state (TS) method, which uses \eqref{eq:SlaterRelation}, to obtain a good approximation to the ionization energy as the HOMO orbital energy of a half-occupied orbital. See Ref.~\cite{Baerends2018JCP} for a recent review of orbital energies and (fractional) occupation numbers. \\
\\
The interest in fractional occupation numbers has stimulated their introduction in density functional theory. Janak's theorem \cite{Janak1978} 
\begin{align}\label{eq:Janak}
\text{Janak:        }\frac{\partial E}{\partial n^{KS}_p} = \epsilon_p^{KS}
\end{align}
expresses relation \eqref{eq:SlaterRelation} for the exact energy and Kohn-Sham orbital energies.  The position of noninteger electron systems (implicit in \eqref{eq:Janak})  in straightforward DFT for single molecules at $T=0$ is actually remarkable if one recognizes that fractional electrons do not exist. One expects that everything that is ``explained" or ``proven" with fractional electrons, can also be treated with integer electron numbers, cf.\ Ref.\ \cite{Gorling2015}. In the case of molecular ionization this is clear: On the one hand Slater's relation \eqref{eq:SlaterRelation} only has meaning if occupation numbers can become fractional, and it can be used to prove Koopmans' theorem in HF using fractional electron numbers (see section \ref{sec:Slater}). It can also be used to calculate with the transition state (TS) method the ionization energies including orbital relaxation from orbital energies at fractional occupation number. On the other hand Koopmans' theorem has first been formulated with just integer electron systems, and Slater's TS method just approximates total energy differences of integer electron systems. Slater's TS method is a mathematical device, it does not lay claim to any physical meaning of the fractional electron numbers. Janak's theorem in exact DFT, however, requires that physical meaning be given to (the energy of) noninteger electron systems (otherwise the derivative of Eq.~\eqref{eq:Janak} would not exist). The introduction of an ensemble description of noninteger electron systems by Perdew, Parr, Levy and Balduz (PPLB) \cite{PerdewParrLevyBalduz1982} is very well known, but it should be kept in mind that this is based on statistical mechanics (using the grand canonical ensemble) \cite{Perdew1985NATO}. It should be realized that he results hold for the average energy as a function of average electron number in for instance a macroscopic gas of molecules in the $T \to 0$ limit, not for a single atom or molecule. \\
\\  
We argue in this paper that Janak's theorem is in fact not a valid theorem in exact density functional theory (meaning DFT for single finite electron systems at $T=0$). Orbital occupation numbers do not properly belong to the KS theory. The KS system of noninteracting electrons will have the KS determinant as wavefunction. Orbitals either occur in that determinant (are occupied) or not. In order for the derivative in \eqref{eq:Janak} to be defined, the exact energy corresponding to a small increase and decrease of some $n_p$  (in a neighborhood of the integer occupation numbers) should be defined. It is not clear what that means in terms of the noninteracting electron system of the KS model (a KS orbital can not ``appear more" or ``appear less" in the determinant). Increase of an orbital occupation above 1, would surely entail increase of the integrated density beyond the integer $N$ value. However, for such densities $E$ is not defined in exact DFT, and therefore the derivative $\partial E/\partial N$ is not defined, and neither would be the derivative with respect to orbital occupation  $\partial E/\partial n_i$.  \\
\\
We discuss in section \ref{sec:Lagrange} the optimization of the density according to the Hohenberg-Kohn theorem, where the constraint of fixed electron number $N$ can in principle be treated with the Lagrange multiplier $\mu=\partial E/\partial N$. The derivative $\partial E/\partial N$, however, is not defined in DFT due to the domain on which the Hohenberg-Kohn functional (and Levy-Lieb and Lieb functionals) are defined being limited to $N$-conserving densities. The standard choice in functional analysis of $E[\rho]=+\infty$ \cite{Lieb1983,vanLeeuwen2003,Lammert2007,Kvaal2014} for $\int\rho d\br \neq N$ is problematic for definition of $\partial E/\partial N$. However, the essential arbitrariness  of $\partial E/\partial N$ does not preclude the optimization of the energy by density variation (at the integer point). It can be done either by
constraining the density variation to normconserving ones, or by the Lagrange multiplier technique.  The theory of constrained derivatives is discussed, in particular how the total functional derivative of the energy $\delta E/\delta \rho(\br)$ can be split into the derivative under the constraint of constant $N$, written $\delta E/\delta_N \rho(\br)$, and the derivative with respect to particle number $N$, $\partial E/\partial N$.   
Only the former is in principle needed, but is difficult to obtain. Therefore a constraint like $\int \rho d\br = N$ is traditionally handled with the Lagrange multiplier technique. The latter, however, requires that  $\partial E/\partial N$ (which enters the formalism as the force of constraint) is known. This raises the question how this technique should be applied if $\partial E/\partial N$ is not defined. Optimization is possible, but requires that a suitable choice of the essentially arbitrary $\partial E/\partial N$ should be made (suitable means finite and continuous in the neighborhood of the integer $N$ point).   \\
\\
In section \ref{sec:Janak} the consequences of $\partial E/\partial n_i$ not being defined are discussed. On the basis of the theory of constrained functional derivatives discussed in section \ref{sec:Lagrange}, a critique of Janak's theorem is given. The conclusion is that Janak's theorem has no validity in (exact) Kohn-Sham DFT. It is also investigated whether a KS like formalism can be set up with possibly fractional occupation numbers (summing up to $N$) by dropping the KS physical model of $N$ noninteracting electrons in $N$ one-electron states in a local potential. This is shown to revert to the KS model.  

Even if systems with a noninteger electron number are not physical, one may still define them as auxiliary systems to obtain useful results for integer electron systems. We have already mentioned Slater's transition state method which uses ionization by half an electron in order to derive an ionization energy (physical quantity) from an orbital energy of an unphysical half-occupied orbital. This method defines, by a specific introduction of (fractional) occupation numbers, the energy  of the auxiliary system as a (nearly) quadratic function of the occupation number. A different introduction of (fractional) occupation numbers, is the one based on statistical mechanical considerations of macroscopic systems at a finite temperature or judiciously chosen $T \to 0$ limit, as introduced by Mermin \cite{Mermin1965} and Perdew et al. \cite{PerdewParrLevyBalduz1982,Perdew1985NATO}. These occupation numbers describe the distribution of the electronic systems in the macroscopic sample over states with different occupations of the one-electron levels, see e.g. the derivation of the Fermi-Dirac distribution in  \cite{AshcroftMermin1976}. So they are physical and describe averages. The search over ensembles of integer electron density matrices of PPLB~\cite{PerdewParrLevyBalduz1982} leads, in the $T \to 0$ limit, to linear interpolation (straight-line behavior) of the average energy and the average electron densities for fractional electron number.

If an analogous procedure is followed for a single finite quantum system (atom or molecule), using an ensemble of density matrices with different electron numbers, density optimization will again result in a specific (``straight'') path through density space and a straight-line energy as a function of fractional electron number, see section \ref{subsec:GrandCanonical}. However, a single system with noninteger electron number is nonphysical. This represents just one out of many possibilities to choose the arbitrary (because nonphysical) $\partial E/\partial N$ in this case.  This choice should not be considered the only viable or even ``exact" definition \cite{KraislerKronik2015,Scheffler2016,LiLuYang2017} of $E[\rho]$ of a single quantum system (atom or molecule)  for $\int \rho d\br \ne N$. The nonexistence of a noninteger electron system makes it impossible to ever obtain a benchmark value for the energy (wavefunction) of such a system.  \\  
It has been argued that the straight-line behavior of the energy can be derived for an atom or molecule from size-consistency arguments, see section \ref{subsec:Locality}. We caution that the underlying assumption is that the functional be \textit{local}, which is not the case for the exact functional. A straight-line energy functional yields dissociation of any diatomic molecule into neutral atoms \cite{PerdewParrLevyBalduz1982} (section \ref{subsec:Dissociation}), which may be considered a strong point. However, we draw attention to the fact that this again relies on the unwarranted assumption of locality of the functional. Indeed, section \ref{subsec:KSdissociation} discusses that the jump of the KS potential when an integer is passed, which is inherent in the straight-line functional, is a local property of a fragment in the dissociation and is not compatible with a correct description of dissociation of a heteronuclear diatomic. \\
\\ 
Finally in section \ref{sec:Slater} the possibility of a  meaningful use of orbital occupation numbers in \textit{approximate} energy expressions is highlighted. The freedom to introduce such additional variables in the total energy expressions is stressed, which may be done in Slater's way or in a different way. The Lagrange multiplier technique can be used to constrain the occupation numbers to physically meaningful values (typically 1 and 0). From this then emerges a physical meaning of the Lagrange multipliers $\{\lambda_i\}$ for this constraint, namely they become equal to the Lagrange multipliers $\{\epsilon_i\}$ for the normalization constraint, $\lambda_i=\epsilon_i$. It is shown that relation \eqref{eq:SlaterRelation} only holds if a particular choice is made for the dependence of the total energy on occupation numbers.  Suppose one uses a model energy that allows to identify a derivative with respect to an occupation number with the related orbital energy according to the Slater relation \eqref{eq:SlaterRelation}, e.g. $\partial E^{model}/\partial n_{LUMO}=\epsilon_{LUMO}$. One may equate the LUMO orbital energy, in this example, to  $(\partial E/\partial N)_+$, but the essential arbitrariness of the latter precludes any conclusion about the value of this orbital energy.  \\   
Section \ref{sec:Summary} summarizes.\\
\\
We should stress that we are dealing here with systems in their ground state at $T=0$. The treatment of finite temperature effects and the statistical physics of large numbers of systems that can exchange energy (canonical ensembles) or also particles (grand canonical ensembles) with a bath are outside the scope of this treatment. The properties of individual quantum systems at $T=0$ are required in the first place, they determine the  behavior of ensembles. Individual systems can be in different states (excited, with different numbers of electrons) but the probabilities of occurrence of such states in statistical ensembles at finite temperatures  is not the subject of Slater's or Janak's relations. \\

%%%%%%%%%%%%%%%%%%%%%%%%%%%%%%
%SECTION sec:Lagrange
%%%%%%%%%%%%%%%%%%%%%%%%%%%%%%%%%%%%%%%%%%%%%%%%%%%%
\section{Energy derivatives and indeterminacy of the Lagrange multiplier for constant electron number $N$}\label{sec:Lagrange}
%%%%%%%%%%%%%%%%%%%%%%%%%%%%%%%%%%%%%%%%%%%%%%%%%%%%

\subsection{The Lagrange multiplier technique and the force of constraint}\label{sec:LagrangeMultiplier}
\begin{figure}
\includegraphics[width=9cm]{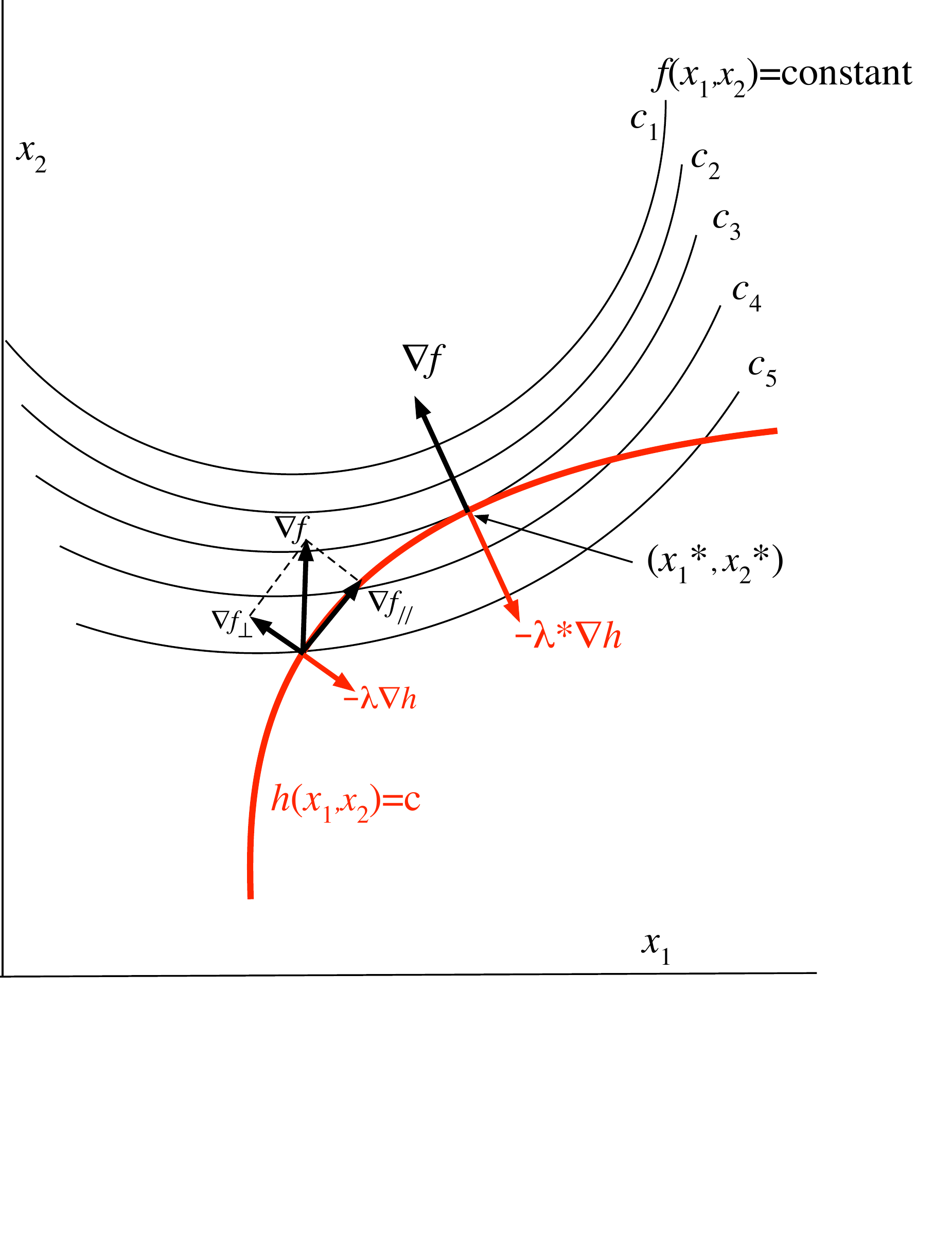}
\caption{A two-dimensional example of application of the Lagrange multiplier: finding the minimum value of the function $f(x_1,x_2)$ under the constraint $h(x_1,x_2)=c$. The crossing point of $h=c$  and $f=c_5$ represents an arbitrary point of the feasible set of points (those obeying the constraint) at which the component $\nabla f_{//}$ in the domain of feasible points is still nonzero. $(x^*_1,x^*_2)$ is the optimal point where $\nabla f_{//}(x^*_1,x^*_2)=0$ and $\nabla f_{\bot}(x^*_1,x^*_2)$ is compensated by the force of constraint $-\lambda^*\nabla h(x^*_1,x^*_2)$. }
\label{fig:LagrangeMultiplier}
\end{figure}

We recall a few salient features of the Lagrange multiplier method \cite{Fletcher1981,NocedalWright1999} for the imposition of constraints when finding an extremum of a function. Fig.\ \ref{fig:LagrangeMultiplier} shows an example in the case of an objective function $f$ (the function for which an extremum has to be found) in two dimensions, under the constraint $h(x_1,x_2)=c$. At an arbitrary point, depicted to the left in the domain of feasible points (those obeying the constraint, i.e.\ on the curve $h(x_1,x_2)=c$), there is a component $\nabla f_{//}$ of $\nabla f$ along the curve of feasible points, which implies that the optimal point has not been reached. The perpendicular component is compensated by the ``force of constraint" exerted by the constraint, $-\lambda \nabla h=-\nabla f_{\bot}$. [It is common practice to use the language of mechanical problems and denote $f$ as a potential in which the minimum is searched and where the probe particle moves in the feasible domain until no forces are any more exerted.] The optimal point $(x^*_1,x^*_2)$ is characterized by the absence of a parallel force, $\nabla f_{//}(x^*_1,x^*_2)=0$, and the Lagrange multiplier $\lambda^*$ at that point is a measure for the strength of the necessary force of constraint.\\
In textbook examples the objective function is typically defined also outside the domain of feasible points,  so that $\nabla f_{\bot}$ is defined. This is also necessary for straightforward application of the Lagrange multiplier technique, which requires all derivatives of the Lagrangean, without constraints, to be zero:
\begin{align}
L(x_1,x_2)&=f(x_1,x_2)-\lambda(h(x_1,x_2)-c) \notag \\
\frac{\partial L}{\partial x_i}&=\frac{\partial f}{\partial x_i}-\lambda\frac{\partial h}{\partial x_i}=0 \quad \forall i
\end{align}
 It may happen of course that the objective function is not defined outside the feasible domain, for instance for physical reasons (as in our case of the Hohenberg-Kohn functional). That is not a problem if $\nabla f_{//}(x_1,x_2)$ can be determined: the point $(x^*_1,x^*_2)$ where $\nabla f_{//}(x^*_1,x^*_2)=0$ will still be the same. Application of the Lagrange multiplier technique, however, requires that $f(x_1,x_2)$ is defined outside the feasible domain, so that the derivative $\nabla f_{\bot}(x_1,x_2)$ can be determined. As a way out one can simply extend in an arbitrary but suitable way the definition of $f$ into the nonfeasible domain, so the derivatives of $f$ exist and $\nabla f_{\bot}$ is defined. The extension is arbitrary in the sense that the actual value of $\nabla f_{\bot}$ is unimportant, it only affects the magnitude of the force of constraint which in this case is only an auxiliary quantity (chosen by the user), not a physical quantity. The extension must be suitable in the sense that the derivative must exist. A nonexistent derivative, for instance different left and right derivatives at the point $(x^*_1,x^*_2)$, precludes the application of the Lagrange multiplier method; the whole purpose of this method is to enable the calculation of the full derivatives, without any possibly very difficult restrictions on those derivatives due to constraints.

%%%%%%%%%%%%%
%%%%%%%%%%%%%%%%%%%%% 

\subsection{Differentiation of $E[\rho]$ and the arbitrariness of $\partial E/\partial N$ and of the Lagrange multiplier for the constraint of fixed electron number}\label{subsec:Differentiation}
%%%%%%%%%%%%%%%%%%%%%%%%%%%%%%%%%%%%%%%%%%%%%%
In the Hohenberg-Kohn theory we have the situation that the objective functional $E_v[\rho]$  is not defined outside the feasible domain of densities that integrate to $N$ electrons, $\int \rho d\br =N$. Part of the derivative $\delta E/\delta \rho$ is therefore undefined (we omit the subscript $v$ when there is no risk of confusion). Following Parr and Bartolotti \cite{ParrBartolotti1983} (see also G\'al \cite{Gal2001}) we make this explicit by writing the density as the product of $N$ and a ``shape function" $\sigma(\br)$ which is normalized to 1:
\begin{align}\label{eq:Nsigma}
\rho(\br) = N \frac{g(\br)}{\int g(\br') d\br'} \equiv N\sigma(\br), \int \sigma(\br)d\br =1
\end{align} 
where $g(\br)$ can be any decent nonnegative function. The density variation $\delta \rho(\br)$ can be split in $N$-conserving and shape-conserving components,
\begin{equation}\label{eq:deltarho}
\delta\rho(\br) =N\delta \sigma(\br) + \sigma(\br)\delta N   \equiv \delta_N \rho(\br) + \delta_\sigma \rho(\br),
\end{equation}
where the subscripts $N$ and $\sigma$ denote variations $\delta \rho(\br)$ where $N$  or $\sigma$ are kept constant, respectively. $\delta \sigma$ should obviously integrate to zero (and should be orthogonal to $\sigma$, $\int \sigma(\br)\delta\sigma(\br)d\br=0$,  see Appendix \ref{sec:derivatives}). The energy variation becomes
\begin{align} \label{eq:deltaE[rho]}
\delta E[\rho] &= \int \frac{\delta E[\rho]}{\delta \rho(\br)}\delta \rho(\br) d\br  \\
&= \int \left[\frac{\delta E[\rho]}{\delta_N \rho(\br)}+\frac{\delta E[\rho]}{\delta_\sigma \rho(\br)} \right] \left( \delta_N \rho(\br)+\delta_\sigma \rho(\br)\right)d\br  \notag  \\
&=\int  \frac{\delta E[\rho]}{\delta_N \rho(\br)}\delta_N \rho(\br) d\br + \int  \frac{\delta E[\rho]}{\delta_\sigma \rho(\br)} \delta_\sigma \rho(\br) d\br \notag 
\end{align} 
where we have introduced the $N$-conserving and shape-conserving constrained derivatives, indicated with subscript $N$ and $\sigma$ in the functional derivative, following G\'al \cite{Gal2001}.  The constrained derivatives for functionals are like partial derivatives for functions, with of course special adaptations.  In going from the second to the third line in \eqref{eq:deltaE[rho]} we have used the fact that the cross terms ($N$-conserving derivative working on shape-conserving density change, and vice versa) are zero. See Appendix \ref{sec:derivatives} for further elaboration and definitions.  Note the analogy with the partial derivatives of functions of more variables, $\delta f = (\partial f/\partial x_1)\delta x_1 + (\partial f/\partial x_2)\delta x_2$, where there are also no cross terms. Now 
\begin{equation}\label{eq:dEdN}
\int  \frac{\delta E[\rho]}{\delta \rho(\br)}\sigma(\br)\delta N d\br = \int  \frac{\delta E[\rho]}{\delta \rho(\br)}\frac{\rho(\br)}{N}d\br \delta N = \frac{\partial E[\rho]}{\partial N}\delta N
\end{equation} 
(cf.\ Eq.\ \eqref{eq:partialEpartialN}) which defines the partial derivative of $E[\rho]$ with respect to $N$, $\partial E/\partial N$. This is $\br$-independent, a constant. However, it is undefined in HK theory, since $E[\rho]$ is not defined for noninteger $N$'s in the neighborhood of an integer $N$.\\
\\ 
The problem of the value of $E_v[\rho]$ outside the domain of normalized densities has been addressed in various ways, usually in discussions of the Hohenberg-Kohn functional $F$, defined according to 
\begin{align}\label{eq:F[rho]}
E_v[\rho]=F[\rho]+\int\rho v d\br; \quad F[\rho]=\brakket{\Psi[\rho]}{\hat{T}+\hat{W}}{\Psi[\rho]}
\end{align}
The expectation values of the kinetic energy $(\hat{T})$ and the electron-electron Coulomb interaction energy $(\hat{W})$ have been defined for fixed electron number ($N$-electron) systems  according to the Hohenberg-Kohn, Levy-Lieb or Lieb prescriptions \cite{DreizlerGross1990,vanLeeuwen2003}. The Lieb functional $F_L[\rho]$, which leads to a convex functional, is always used in mathematical treatments (see earlier work by Levy \cite{Levy1979} and Valone \cite{Valone1980} on searches over density matrices of $N$-electron states) 
\begin{align}\label{eq:F_L}
&\text{Lieb}: \quad F_L[\rho]=\inf_{\hat{D}\to \rho} Tr\hat{D}\left(\hat{T}+\hat{W}\right) \\
& \hat{D}=\sum_{i=1}^n \lambda_i \DM{\Psi_i^N}{\Psi_i^N}; \;  0 \le \lambda_i \le 1,  \sum_{i=1}^n \lambda_i=1 \notag  \\
\end{align}
Lieb \cite{Lieb1983} in his study of functional analysis aspects of DFT proposed to define $F_L[\rho]$ as $F_L[\rho]=+\infty$ for densities that fall outside the domain of positive functions that integrate to $N$. Infinite values are well-defined in the theory of convex functionals and they are usually introduced to deal in a simple way with domain issues \cite{vanLeeuwen2003}. This choice has usually been followed in analyses of the differentiability of $F_L$ \cite{Eschrig1996,vanLeeuwen2003,Lammert2007,Kvaal2014,GiesbertzRuggenthaler2018}. It is however very different from the choice of Ref.~\cite{PerdewParrLevyBalduz1982} to define $F_L[\rho]$ for such densities by an extension of the density matrices over which the search is performed to ensembles of density matrices with different integer $N$ values, see section \ref{subsec:GrandCanonical}. That search then leads to linear interpolation of the energy between the $N$- and $(N+1)$-electron ground state energies at the electron-rich side (with derivative $-A$) and between the $N$- and $(N-1)$-electron ground state  energies at the electron deficient side (with derivative $-I$). But an infinite derivative or the lack of differentiability at the integer point (different left and right derivatives) invalidates the Euler-Lagrange equation \eqref{eq:EulerLagrange} (the force of constraint cannot be determined). Lieb has also indicated that more regular choices can be made, such as proportionality of $F_L$ to $\int \rho d\br=N$ ($N$ noninteger), e.g. $F_L[\rho]=(\int\rho d\br)F_L[\sigma(\br)]$ with straightforward derivative (see remarks at theorem 3.3  \cite{Lieb1983}). We stress here the arbitrariness of the definition of $F_L$ in the noninteger $N$ domain, hence the freedom to make a suitable choice.\\      
In applications of the Lagrange multiplier method it is required that  $\partial E[\rho]/\partial N$ is a defined quantity. Then we should make sure our results are independent of the choice we have made.  In principle (when trying to determine the $N$-electron ground state density $\rho_0^N$) we do not need $\partial E[\rho]/\partial N$ because we could only search in the domain of $N$-electron densities,  so at  $\rho_0^N$ we must have, using the density shape factor $g$ of Eq.\ \eqref{eq:Nsigma} (cf.\ \cite{ParrBartolotti1983,ParrYang1989})
\begin{equation}\label{eq:dEdg1}
\frac{\delta E[\rho_0^N]}{\delta g(\br)} =\int  \frac{\delta E[\rho_0^N]}{\delta \rho(\br')} \frac{\delta \rho(\br')}{\delta g(\br)}d\br'=0
\end{equation}
where we have used the chain rule. Since \cite{ParrYang1989}
\begin{equation}\label{eq:drhodg}
\frac{\delta \rho(\br')}{\delta g(\br)}=\frac{N}{\int g(\br'')d\br''}\left[ \delta(\br-\br')-\frac{g(\br')}{\int g(\br'')d\br''} \right]
\end{equation}
we have from \eqref{eq:dEdg1} and \eqref{eq:drhodg} at $\rho_0^N$
\begin{equation}\label{eq:dErho0drho}
 \frac{\delta E[\rho_0^N]}{\delta \rho(\br)}=\int  \frac{\delta E[\rho_0^N]}{\delta \rho(\br')}\frac{g(\br')}{\int g(\br'')d\br''}d\br'
\end{equation}
This shows that at $\rho_0^N$ the functional derivative $\delta E/\delta \rho(\br)$ is not a function of $\br$ but is a constant. This fits in with the fact that at $\rho_0^N$ the functional derivative for $N$-conserving variations of $\rho$, $\delta E[\rho_0^N]/\delta_N \rho(\br)$, is zero, see next paragraph (and in agreement with Eq.~\eqref{eq:dEdg1}). So the full derivative $\delta E[\rho_0^N]/\delta \rho(\br)$ reduces to the constant (independent of $\br$) $\partial E[\rho_0^N]/\partial N$, as is evident by comparing \eqref{eq:dErho0drho} with \eqref{eq:dEdN}. 
We cannot derive this constant from Eq.\ \eqref{eq:dErho0drho}, since any constant value is compatible with that equation. That is in order, since  we have required the functional derivative to be zero only with respect to shape variation, Eq.\ \eqref{eq:dEdg1}.  We can expect to obtain a result that is compatible with any value of $\partial E[\rho_0^N] / \partial N$. The fact that $\partial E[\rho_0^N] / \partial N$ is not defined in DFT, since HK theory does not provide values of the functional $E[\rho]$ for $N$-nonconserving $\rho$, and therefore not of the derivative, is not a problem. It should suffice to only require the derivative to be zero for $N$-conserving variations of $\rho$, $\delta E[\rho_0^N]/\delta_N \rho(\br)=0$, or $\delta E[\rho_0^N]/\delta g(\br)=0$. There is a risk that it is not recognized that the undefined constant is part of the total derivative, but not of the constrained derivative, see comments on Eqn.~\eqref{eq:constant} and see Appendix \ref{sec:ParrYang}.   \\
\\
In practice constrained derivatives are not used since they are hard or impossible to obtain, and the standard Lagrange multiplier method is used because it has the great advantage that it allows to work with unconstrained derivatives, which are usually straightforwardly obtained. We try to find the minimizing density from
\begin{equation}\label{eq:EulerLagrange}
\int  \frac{\delta }{\delta \rho(\br)} \left[ E[\rho] -\mu\left(\int \rho(\br)d\br - N\right)\right]\delta \rho(\br) d\br=0
\end{equation}
($N$ is integer). This should hold for arbitrary $\delta \rho$. We may first choose for $\delta \rho(\br)$ just $\delta_\sigma \rho(\br)$ and must have at the optimum density $\rho_0^N$ 
\begin{equation}\label{eq:EulerLagrange2}
\int \left[ \frac{\delta E[\rho_0^N] }{\delta_N \rho(\br)} +  \frac{\partial E[\rho_0^N]}{\partial N} - \mu \right]\delta_\sigma\rho(\br)d\br=0.
\end{equation}
Since  $\int [\delta E[\rho]/\delta_N\rho(\br)]\delta_\sigma\rho(\br)=0$ (see appendix \ref{sec:derivatives}) this yields $\mu^*= \partial E[\rho_0^N]/\partial N$: the force of constraint is determined by the choice of  $\partial E/\partial N$. Using next the remaining $\delta\rho(\br)$ space of norm-conserving $\{\delta_N\rho(\br)\}$, and substituting $\mu^*= \partial E[\rho_0^N]/\partial N$ it is clear that at $\rho_0^N$ we must have
\begin{align}\label{eq:dErho0Nd_Nrho}
&\int \frac{\delta E[\rho_0^N]}{\delta_N \rho(\br)} \delta_N\rho(\br) d\br=0 \\
&\to  \int \frac{\delta E[\rho_0^N]}{\delta_N \rho(\br)} \delta\rho(\br) d\br=0 \to \frac{\delta E[\rho_0^N] }{\delta_N \rho(\br)}=0 \notag 
\end{align}
where the second line follows since we already had $\int \delta E[\rho_0^N]/\delta_N \rho(\br) \delta_\sigma\rho(\br) d\br=0$, as used above. Here it does not suffice to use only the  subspace $\{\delta_N\rho(\br)\}$ and require that the $N$-conserving derivative is only zero for application on that subspace. That requirement alone does not completely fix $\delta E[\rho_0^N]/\delta_N \rho(\br)$.\\
There is a pitfall lurking here.  It has often been noted in the literature \cite{ParrBartolotti1983,PerdewLevy1983,ParrYang1989,ParrLiu1997,vanLeeuwen2003,
Gorling2015} that, since only norm conserving density variations are allowed in HK theory, derivatives like $\delta E/\delta \rho(\br)$ (or $\delta F[\rho]/\delta \rho(\br)$) are only defined up to a constant: if $\int \delta_N\rho(\br)d\br=0$, then 
\begin{align}\label{eq:constant}
\delta E &= \int d\br \frac{\delta E}{\delta \rho(\br)} \delta_N\rho(\br) \notag \\
\text{is equal to }\delta E &= \int d\br \left(\frac{\delta E}{\delta \rho(\br)} +C \right)\delta_N\rho(\br)
\end{align}
It then appears that the same argument can be applied to the constrained derivative $\delta E/\delta_N \rho(\br)$, which is then also believed to be only defined up to a constant. However, full definition of the functional derivatives requires their ``operation" to be defined on arbitrary $\delta \rho$, not only normconserving $\delta_N\rho$. As we have seen, this prohibits a free constant in  $\delta E[\rho_0^N]/\delta_N \rho(\br)$. In the Hohenberg-Kohn theory there $is$ such an undefined constant in the $full$ derivative $\delta E[\rho_0^N]/\delta \rho(\br)$, which stems from the HK restriction to integer electron systems, causing $\partial E/\partial N$ to be an undefined quantity in the theory. A constant is not allowed in the constrained derivative $\delta E[\rho_0^N]/\delta_N \rho(\br)$ since it would give $\delta E[\rho_0^N]/\delta_N \rho(\br)$ a component in the subspace $\{\delta_\sigma \rho(\br)\}$, to which it should be  ``orthogonal" (note the analogy of $\delta E[\rho_0^N]/\delta_N \rho(\br)$ with $\nabla f_{//}$ of Fig.~\ref{fig:LagrangeMultiplier}). See also discussion in Appendix \ref{sec:ParrYang}.\\
\\
We have seen that it is advisable to define $E[N\sigma]$ as a continuous function of $N$ in a neighborhood of the integer value $N$, so that the derivative $\partial E[N\sigma] / \partial N$ exists. We mentioned such a choice, derived from Lieb, earlier. Parr and Yang (\cite{ParrYang1989}, p. 84) suggest to take a parabolic fit over the three points $E(N-1), E(N)$ and $E(N+1)$. This yields a continuous $E(N)$ in the neighborhood of $N$, with $\partial E[\rho_0^N]/ \partial N=-(I+A)/2$, i.e.\ Mulliken's electronegativity.  G\"orling \cite{Gorling2015} has analyzed the arbitrary constant in the total derivative of the energy (or equivalently the Hohenberg-Kohn functional $F[\rho]$) along similar lines, and fixes it by adhering to the energy behavior at noninteger $N$ according to the PPLB straight lines picture of Ref.\ \cite{PerdewParrLevyBalduz1982}. This has the difficulty of different left and right derivatives at the integer point. It would also be possible to choose $E[\rho]$ independent of $N$ in the neighborhood of $N$, which would yield $\partial E[\rho_0^N]/ \partial N= 0$. All of this is compatible with the fact that  $\partial E/\partial N$ is arbitrary. \\

%%%%%%%%%%%%%%%%%%% Section Janak's Theorem %%%%%%%%%%%%
 \section{Janak's theorem: not a DFT theorem}\label{sec:Janak}
%%%%%%%%%%%%%%%%%%%%%%%%%%%%%%%%%%%%%%
The relation \eqref{eq:Janak} is widely quoted as Janak's theorem and is often considered a fundamental relation in Kohn-Sham DFT. However, there are difficulties with this relation. One can try to formulate relation \eqref{eq:Janak} for a mathematically defined energy where occupation numbers (that supposedly can vary continuously) are introduced. Janak \cite{Janak1978} introduced  occupation numbers in a total energy expression (called $\tilde{E}$), which we call $E^{KS1}$ because it has occupation numbers to the first power (see section \ref{sec:Slater} for an alternative), 
\begin{align}\label{eq:EKS1}
&E^{KS1}=\sum_{\sigma=\alpha, \beta} \sum_p n_p^\sigma \brakket{\phi_p^\sigma}{-\frac{1}{2}\nabla^2+v_{ext}}{\phi_p^\sigma} \notag \\
&+\frac{1}{2} \sum_{\substack{p,q\\\sigma, \tau=\alpha,\beta}}  n^\sigma_p n^\tau_q  \int \frac{\phi_p^{\sigma\star}(\bx_1)\phi_q^{\tau\star}(\bx_2)\phi_p^{\sigma}(\bx_1)\phi_q^{\tau}(\bx_2)}{|\br_1-\br_2|}  d\bx_1 d\bx_2  \notag \\
&+ E_{xc}[\rho^\alpha,\rho^\beta], \\
&\rho^\sigma(\br)=\sum_p n^\sigma_p |\phi^\sigma_p(\br)|^2   \notag
\end{align}
The linear dependency on occupation numbers in the expression for the density and for the one-electron terms, which had earlier been introduced in the X$\alpha$ and Hartree-Fock total energy expressions, is arbitrary (cf.\ section \ref{sec:Slater}) but plausible. For integer values (1 and 0) of the occupation numbers there is no effect. In order for $E^{KS1}$ to have a defined derivative with respect to an occupation number, it is necessary that $E_{xc}[\rho^\alpha,\rho^\beta]$ is defined for densities with fractional electron numbers (a neighborhood of an integer $n_p$).  But there is a problem with the value of $E_{xc}[\rho]$ in Eq.~\eqref{eq:EKS1} at noninteger electron numbers. Eq.~\eqref{eq:EKS1} is in fact the defining equation for $E_{xc}[\rho]$, its value is determined by all other terms (in particular $E^{KS1}[\rho]$) which need to be defined. But at noninteger electron numbers the energy $E^{KS1}[\rho]$ (which should be $E_v[\rho]$) is not defined, and therefore $E_{xc}[\rho]$ is not defined. There are simply no physical systems with a noninteger number of electrons, for which an energy $E_v[\rho]$ would be obtainable. Also the Kohn-Sham construction of a noninteracting system of electrons is not possible for fractional electron numbers, so $E_{xc}[\rho]$ at noninteger electron densities is not a physical quantity. It is of course possible to choose an energy for noninteger electron densities in some way or another, for instance using ensembles of integer electron systems (see section \ref{subsec:GrandCanonical}), but then it has to be kept in mind that no physical information can follow from such a man-made choice. Janak \cite{Janak1978} did not specify $E^{KS1}[\rho]$ for noninteger densities. \\
We may wonder if Janak's theorem \eqref{eq:Janak} may at least have meaning at the integer point, with occupation numbers introduced according to $E^{KS1}$ \eqref{eq:EKS1}.  Let us consider explicitly the derivative
\begin{align}\label{eq:Janakderiv}
&\left. \frac{\partial E^{KS1}}{\partial n^\sigma_p}\right|_{\rho_0^N} = \brakket{\phi^\sigma_p} {-\frac{1}{2}\nabla^2+v_{ext}+v_{Coul}} {\phi^\sigma_p}  \notag \\
&\qquad \qquad \qquad  + \int  \left. \frac{\delta E_{xc}}{\delta \rho(\br)}\right|_{\rho_0^N}  \left. \frac{\partial \rho}{\partial n_p^\sigma}\right|_{\rho_0^N} d\br  \\
&=\brakket{\phi^\sigma_p} {-\frac{1}{2}\nabla^2+v_{ext}+v_{Coul}+(v_{xc}+C)} {\phi^\sigma_p}=\epsilon^\sigma_p+C \notag
\end{align} 
(we consider for simplicity the spin-compensated case where $v_{xc}$ is the same for spin-up and spin-down orbitals). We have indicated the occurrence of an unknown constant $C=\partial E_{xc}/\partial N$. It is unknown because the derivative for a change $\delta N$ in the number of electrons is unknown, see section \ref{sec:Lagrange}.  In the 
total derivative of $E_{xc}$,
\begin{align}
 \frac{\delta E_{xc}[\rho_0^N]}{\delta \rho(\br)} =  \frac{\delta E_{xc}[\rho_0^N]}{\delta_N\rho(\br)}  + \frac{\partial E_{xc}[\rho_0^N]}{\partial N}, 
 \end{align}
the $N$-conserving derivative $\delta E_{xc}[\rho_0^N] / \delta_N \rho(\br)$ taken at the $\rho_0^N$ point may be identified with the exchange-correlation potential $v_{xc}$ (see below), but still there is the undefined constant $\partial E_{xc}/\partial N$.  So the derivative $\partial E / \partial n_p$ is not a defined quantity in Kohn-Sham DFT, not even at the ground state density. It certainly is not defined at fractional electron densities, where not even the $\tilde{N}$-conserving derivative $\delta E[\rho]/\delta_{\tilde{N}}\rho(\br), \int \rho(\br) d\br = \tilde{N}=N+\omega$ is defined. At the $N$-electron ground state density $\rho_0^N$ we are free to choose the constant $\partial E_{xc}[\rho_0^N]/\partial N$, but then we can not draw conclusions about the physics of a system from that choice. \\
 \\

The undetermined constant in the functional derivative, i.e.\ in the xc potential, also arises in the derivation of the Kohn-Sham equations, but it is harmless there. In the traditional derivation of the KS equations \cite{KohnSham1965a} one  minimizes the KS expression for the energy with respect to variations in the density via the orbitals, from which the one-electron Kohn-Sham equations follow. The application of the chain rule in the variation of the $E_{xc}$ term in the energy leads to the exchange-correlation potential $v_{xc}(\br)=\delta E_{xc}/\delta \rho(\br)$, without any precaution being taken that the total derivative is not defined (i.e.\  not for $N$-nonconserving density variation). However, the chain rule for the $E_{xc}$ term gives  
\begin{align}
\delta E_{xc}&= \sum_p^N \int \frac{\delta E_{xc}}{\delta \phi_p(\br)} \delta \phi_p(\br) d\br \notag  \\
&=\sum_p^N \int \int \frac{\delta E_{xc}}{\delta \rho(\br')}\frac{\delta \rho(\br')}{\delta \phi_p(\br)}d\br'  \delta \phi_p(\br) d\br \notag \\
&=\sum_p^N \int \left(\frac{\delta E_{xc}}{\delta_N\rho(\br)}+\frac{\partial E_{xc}}{\partial N}\right)\phi^*_p(\br)\delta\phi_p(\br) d\br\notag \\
&=\sum_p^N \int (v_{xc}(\br)+C) \phi_p^*(\br)\delta \phi_p(\br) d\br
\end{align}
where $v_{xc}(\br)$ is the defined derivative $\delta E_{xc}/\delta_N\rho(\br)$ (because $E_{xc}$ is defined for $N$-conserving densities) and one is left with the undefined constant $C=\partial E_{xc}/\partial N$. But  in this case this arbitrary constant in the KS potential does not have any physical consequences. A constant in the potential (which extends over all space, including asymptotic regions) just shifts the whole eigenvalue spectrum up or down. This is the well known gauge freedom of a local potential, which can be eliminated by always choosing the potential to go to zero at infinity, so that all calculations work with the same gauge and the orbital energies become comparable. This is commonly done. Although the commonly used definition $v_{xc}(\br)=\delta E_{xc}/\delta \rho(\br)$ is strictly speaking not correct, or not complete, since it ignores the fact that the HK theorem only allows $N$-conserving densities, we now see that this omission does not have any consequences. There is not a similar saving grace in the derivation of Janak's theorem, since there the undefined constant exactly expresses that it is undefined what the theorem claims to tell, namely how the energy changes under a change $\delta N=\delta n_p$.    \\
\\  
One may wonder if fractional occupations could be meaningfully introduced by abandoning the Kohn-Sham model system of $N$ noninteracting electrons in a local potential. Parr and Yang (\cite{ParrYang1989}, \S 7.6) have raised the question if then the Janak theorem could be put on a secure footing. They consider a generalization of the noninteracting kinetic energy $T_s[\rho]$ to a Janak kinetic energy
\begin{align}\label{eq:T_J}
T_J[\rho]=\min_{n_i, \psi_i \to \rho} \sum_i^\infty n_i \brakket{\psi_i}{-\frac{1}{2}\nabla^2}{\psi_i}
\end{align}
where the search is over all possible $n_i$ $(0 \le n_i \le 1)$ and orthonormal orbitals $\psi_i$ yielding the given density (constraining it to $N$ electrons) 
\begin{align}
\rho(\br)=\sum_i^\infty n_i  |\psi_i(\br)|^2 \qquad \sum_i n_i = N
\end{align} 
The exact total energy functional can now be written
\begin{align}\label{eq:EJanak}
E[\rho]=T_J[\rho]+\int v(\br)\rho(\br)d\br + W_{Hartree} + E^J_{xc}[\rho]
\end{align}
where $W_{Hartree}=(1/2)\int \rho(\br_1)\rho(\br_2/r_{12}d\br_1 d\br_2$ and $E^J_{xc}[\rho]$ is defined by this equation and is different from the standard $E_{xc}[\rho]$ if $T_J[\rho]$ would be different from $T_s[\rho]$.  
The question is if this set-up might lead to noninteger $n_i$. Then  occupation numbers would have been introduced, possibly fractional, without the physical model of Kohn and Sham of noninteracting electrons in a local potential, which is defined with integer (1 and 0) occupation numbers. With integer $n_i$, the model of Eqns \eqref{eq:T_J} and \eqref{eq:EJanak} would reduce to the standard noninteracting kinetic energy of the KS determinantal wavefunction with the given $\rho$ and lowest kinetic energy. \\
\\
For $E[\rho]$ of \eqref{eq:EJanak} to be defined,  $\rho$ must be a density belonging to a ground state wavefunction (HK) or at least an $N$-electron wavefunction or density matrix (Levy-Lieb). Then $E^J_{xc}[\rho]$ is only defined for such integer-$N$ densities.   
Parr and Yang \cite{ParrYang1989} derive, at a given set of $\{n_i\}$, the  KS-like equations for the $\psi_i$ which have $n_i \ne 0$
\begin{align}\label{eq:KSJeq}
\left[ -\frac{1}{2}\nabla^2 +v^J_{eff}(\br)\right] \psi_i= \epsilon_i \psi_i
\end{align}
where $v_{eff}^J=v(\br)+v_{Hartree}(\br)+v_{xc}^J(\br)$ with $v_{xc}^J(\br)=\delta E_{xc}^J/\delta \rho(\br)$.
Differentiating the total energy with respect to an occupation number $n_i$ is again problematic since the energy $E[\rho+\delta\rho]$ is not defined for a small $N$-nonconserving density change $\delta \rho=\delta n_i|\psi_i|^2$, and neither is $E_{xc}^J[\rho+\delta\rho]$. Ignoring this problem and again applying the chain rule for the derivative of $E_{xc}^J$ yields 
\begin{align}\label{eq:EJ-derivative}
\frac{\partial E}{\partial n_i} = \epsilon_i
\end{align}
The behavior of the energy \eqref{eq:EJanak} under occupation number changes according to Eq.~\eqref{eq:EJ-derivative} is well known to lead to Aufbau, i.e.\ $n_i=1$ for the lowest orbitals emerging from the optimization, and $n_a=0$ for the remaining ``virtual" orbitals. This is easily seen by considering infinitesimal occupation number changes. The Janak-Kohn-Sham system with $T_J[\rho]$ instead of $T_s[\rho]$ then simply reverts to the Kohn-Sham system. \\

This raises the question if the Janak kinetic energy is actually different from $T_s[\rho]$. We can prove that this is not the case, i.e.\ the orbitals resulting from the minimization of Eq.~\eqref{eq:T_J} are the KS orbitals belonging to $\rho$ and the optimal occupation numbers obey Aufbau, so
\begin{equation}\label{eq:T_J=T_S}
T_J[\rho]=T_s[\rho].
\end{equation}
Let us minimize $T_J[\rho]$ by varying $\{n_i\}$ and $\{\psi_i\}$ under the constraint that $\sum_i^\infty n_i|\psi_i(\br)|^2=\rho(\br)$ at each $\br$, for which purpose we introduce the $\br$-dependent Lagrange multiplier $\mu(\br)$. Normalization of the $\psi_i$ is maintained with the usual Lagrange multipliers $\epsilon_i'$. The Lagrangean $L_J$ becomes
\begin{align}\label{eq:LagrangeanT_J}
L_J=&\sum_i^\infty n_i \brakket{\psi_i}{-\frac{1}{2}\nabla^2}{\psi_i} \notag \\
&+\int \mu(\br)\left(\sum_i^\infty n_i|\psi_i(\br)|^2-\rho(\br)\right)d\br \notag \\
&-\sum_i^\infty \epsilon_i'\left(\int |\psi_i(\br)|^2d\br-1 \right) 
\end{align}    
For a fixed set of $\{n_i\}$, optimization of the orbitals \cite{ParrYang1989} yields the equations
\begin{align}\label{eq:KSeqJ}
 \left( -\frac{1}{2}\nabla^2 +\mu(\br) \right) \psi_i = \epsilon_i \psi_i 
\end{align}
for the occupied orbitals, where $\epsilon_i=\epsilon_i'/n_i$. The Lagrange multiplier $\mu(\br)$ acts as local potential in this one-electron equation. Orthogonality of the orbitals then follows from the hermiticity of the operator and need not be enforced separately \cite{ParrYang1989}. Before determining $\mu(\br)$ in the usual way from the constraints, we note that with a fixed potential $\mu(\br)$ the energy of the noninteracting electron system with Janak kinetic energy would be
\begin{align}\label{eq:E_s^J}
E_s^J=\sum_i^\infty n_i \brakket{\psi_i}{-\frac{1}{2}\nabla^2}{\psi_i} +\int \mu(\br)\rho(\br) d\br
\end{align}
Minimization of this energy by variation of the $\{n_i\}$ and $\{\psi_i\}$ under the constraint $\sum_i^\infty n_i|\psi_i(\br)|^2=\rho(\br)$ leaves the $\int \mu(\br)\rho(\br)d\br$ term invariant and is therefore equivalent to (should lead to the same $\{n_i\}$ and $\{\psi_i\}$ as) the minimization in the definition of Janak kinetic energy according to \eqref{eq:T_J}. Considering then variation of the $\{n_i\}$, we note that now there is no problem with taking the derivative with respect to $n_i$ in \eqref{eq:E_s^J} since an exact energy for a noninteger electron system is not required for its definition. The derivative will just be
\begin{equation}
\frac{\partial E_s^J}{\partial n_i} = \int \psi_i^*(\br)\left( -\frac{1}{2}\nabla^2 + \mu(\br) \right) \psi_i(\br) d\br = \epsilon_i
\end{equation}
So the minimum will be obtained for Aufbau. If the fixed set of $\{n_i\}$ with which we started was inadvertently not the Aufbau choice of $\{n_i\}$, which we may call $\{n_i^A\}$, the condition $\sum_i n_i^A|\psi_i(\br)|^2=\rho(\br)$ will not be obeyed with the present set of $\{\psi_i\}$ and $\{n_i\}$. So then we have to repeat the process with occupation numbers according to Aufbau. The potential $\mu^A(\br)$ will become exactly the KS potential $v_s[\rho](\br)$ belonging to density $\rho(\br)$, since that KS potential is unique.  So $T_J[\rho]$ is not a new ``Janak" kinetic energy but it is just $T_s[\rho]$. There is no Janak kinetic energy and no Janak-Kohn-Sham model.\\
\\
Recently Li et al.~\cite{LiLuYang2017} studied the Janak construction with a different purpose, namely to see if in case of a noninteger $\rho$ this would not lead to occupation of higher virtual KS orbitals but just to the fractional occupation of frontier orbitals (Aufbau). They concluded to this Aufbau behavior of the occupation numbers. However, we have indicated that $E[\rho]$ is problematic in case $\rho$ is a noninteger density.  
Even if a $T_J[\rho]$  exists for noninteger densities, it can not be used in Eq.~\eqref{eq:EJanak}, since with a noninteger density,  $E[\rho]$ in that equation is not defined.  \\
\\
Valiev and Fernando \cite{ValievFernando1995} have objected against the Janak theorem on somewhat different grounds. They point out that the results of the work by Englisch and Englisch \cite{EnglischEnglisch1984b} imply that the energy of a noninteracting system at fractional values of the occupation numbers (except for the degenerate levels at the Fermi energy) would not be differentiable with respect to the density.\\
\\
We are not considering here fractional occupations of degenerate levels at the Fermi energy, which is theoretically \cite{Levy1982,Lieb1983} and practically \cite{SchipperGritsenko1998,UllrichKohn2001} a well understood case, and neither do we consider density matrix functional theory (DMFT). Using the spectral resolution of the 1-matrix $\gamma(\bx_1,\bx_1')=\sum_p n_p \psi_p(\bx_1)\psi^*_p(\bx_1')$ the optimization in DMFT is with respect to both the natural orbitals and the occupation numbers. In DMFT the latter are an intrinsic part of the theory, they typically become all fractional.  The inequality constraints $0\le n_p \le 1$ have to be applied with the Karush-Kuhn-Tucker method, see Giesbertz and Baerends \cite{GiesbertzBaerends2010}. Orbital energies start to play a very different role, cf.\ Gilbert's \cite{Gilbert1975} famous finding of degeneracy for all fractionally occupied natural orbitals, which usually means all natural orbitals.

%%%%%%%%%%%%%%%%%%%%%%%%%%%%%%%%%%%%%%%%%%%%%%%%%%
%%%%%%%%%%%%%%%%%%%%%%%%%%%%%%%%%%%%%%%%
\section{Energy behavior at noninteger electron numbers. }\label{sec:Ensemble}

Ever since the seminal PPLB paper \cite{PerdewParrLevyBalduz1982}, the straight-line behavior of the energy over the $(N,N+1)$ and $(N,N-1)$ intervals has received much attention in the literature. It should be emphasized that PPLB addresses the issue of fluctuating particle number as can occur in macroscopic samples at finite temperature,  and use the grand canonical ensemble of statistical mechanics to treat this variable particle number. They derive straight-line behavior in the $T \to 0$ limit (see the full details treated in \cite{Perdew1985NATO}). On the basis of this work it is sometimes assumed that even for a single quantum system (atom or molecule)  $\partial E/\partial N$ is NOT arbitrary, but has to be $-I$ on the $(N,N-1)$ interval and $-A$ on the $(N,N+1)$ interval, with discontinuous derivative at $N$.  We do not feel that it is a correct interpretation of the results of Ref.\ \cite{PerdewParrLevyBalduz1982} to consider this behavior as mandatory or ``exact" for \textit{a single molecule}. For the common case of DFT calculation on a single molecule at $T=0$ the interpolation of the density between the integer $N$ and $N \pm 1$ ground state densities and the corresponding straight-line behavior of the energy constitute just one choice for the  essentially arbitrary continuation (see section \ref{sec:Lagrange}) to nonphysical fractional electron number.  We will discuss this in subsection  \ref{subsec:GrandCanonical}. \\
 One can also argue in favor of linear energy behavior on the basis of size-consistency requirements for dissociation of molecules, as has been done by Yang, Zhang and Ayers \cite{YangZhangAyersPRL2000,Ayers2008fractional}. This is the subject of section \ref{subsec:Locality}, where we argue that it is not size-consistency but the (unwarranted) requirement of locality of the functional that leads to the linear behavior. In section \ref{subsec:Dissociation} we recall that application of the straight-line energy (assuming it holds for single quantum systems) has the well-known success of describing dissociation of a molecule into integer electron fragments, but we caution that again the unwarranted local approximation is invoked. In section \ref{subsec:KSdissociation} it is shown that the local approximation, which is inherent in the application of the straight-line energy to dissociation, also causes it to fail: the derivative discontinuity jump in the KS potential is not quantitatively correct, in that it does not lead to proper dissociation of a heterogeneous electron pair bond. \\
 
 %%%%%%%%%%%%%%%%%%%%%%%%%%%%%%%%%%%%%%%%%%%%%%%% 
\subsection{Straight-line energy behavior from grand canonical ensemble considerations and for single molecules.}\label{subsec:GrandCanonical}
%%%%%%%%%%%%%%%%%%%%%%%%%%%%%%%%%%%%%%%%%%%%%%%%% 
 
The statistical mechanical approach of \cite{PerdewParrLevyBalduz1982,Perdew1985NATO} implies that the straight-line behavior applies to the \textit{average} energy of the molecules in a dilute macroscopic gas of molecules, that can exchange electrons with a reservoir, at the properly taken $T \to 0$ limit. Also a very simple consideration at just $T=0$ makes that clear. We can then take $\mu$ as a tunable parameter of the reservoir, regulating the energy involved in electron transfer  to and from the reservoir. A physical realization is discussed by Perdew in  \cite{Perdew1985NATO} taking the reservoir to be a metal with workfunction $\Phi$ having negligible coupling integrals with a far away molecule.  Then at $T=0$ when $\mu=-\Phi$ drops below minus the molecular ionization energy, evidently the ground state energy corresponds to all molecules giving up an electron to the reservoir, i.e.\ become ionized. Similarly, as soon as $\mu$ would rise above minus the electron affinity, all molecules would turn into negative ions. At  the points $\mu=-I$ and $\mu=-A$ the number of electrons on the molecules is undetermined. This is also what Perdew observes (see Eq.\ (27) of \cite{Perdew1985NATO}) for a molecule and a metal with work function $\Phi$ as reservoir. At those specific values of $\mu$ many situations are possible, each characterized by specific probabilities to find $N$ and $N-1$  electrons on a molecule (for $\mu=-I$) or $N$ and $N+1$ electrons (for $\mu=-A$). Ref.\ \cite{Perdew1985NATO} discusses the wavefunctions one may construct for the reservoir-molecule system that yield specific probabilities for the possible integer number of electrons on a molecule. When there is negligible coupling between molecule and reservoir, the wavefunction construction can yield the same probabilities as a mixture of density matrices  \cite{PerdewParrLevyBalduz1982,Perdew1985NATO}. This yields a clear interpretation of the noninteger electron number and the corresponding energy as the averages of these quantities over all molecules of the dilute gas. For instance, if  at $\mu=-A$ $m$ electrons (say) are taken up from the reservoir by the gas with a very large number $M$ of molecules, they have to go (at zero temperature) to $m$ molecules, the change of the average number of electrons per molecule is $\Delta\overline{N}=m/M$ and the average energy change is $\Delta \overline{E}= -mA/M$.  One obtains the derivative of the average energy  $\partial \overline{E} / \partial \overline{N} = \Delta \overline{E}/\Delta \overline{N}= (-mA/M)/(m/M)=-A$ at $\overline{N}=N+\omega$ electrons. Going to $\mu=-I$ and $\overline{N}=N-\omega$ electrons, the derivative makes a quantum jump at integer $N$ to $-I$ because of the quantum nature of the molecules. At integer $N$ there is not a defined derivative $\partial \overline{E}/\partial \overline{N}$ (because these consideratioins are at $T=0$ and do not take into account the proper $T \to 0$ limit). It is to be emphasized that this behavior of the average energy and its derivative with respect to the average number of electrons $\overline{N}$ do not imply the same properties for any individual atom or molecule.  No actual system can have a fractional number of electrons. Such systems are fictitious and we cannot calculate a wavefunction for them or get any information such as the energy for them. \\
\\
The arbitrariness of the energy of a single atom or molecule at noninteger density implies that one may construct a functional which has some convenient behavior. It is important then, of course, to refrain from conclusions about the physical (integer electron) system from the arbitrary choice. For instance, a straight-line behavior of the energy for a single molecule at noninteger electron number is obtained along similar lines as in Ref.\ \cite{PerdewParrLevyBalduz1982}, see below, but now we relinquish any statistical mechanical underpinning and are dealing frankly with the nonphysical system of a single molecule with fractional electron number. We first recall that following work by Levy \cite{Levy1979}  and Valone \cite{Valone1980a}, Lieb \cite {Lieb1983} formulated the Lieb functional  in terms of a constrained search over an ensemble of $N$-electron pure state density matrices, Eq.~\eqref{eq:F_L}. The Lieb functional has the important property of being convex.
The further step can be taken of considering a density that integrates to a noninteger number of electrons, where the choice is made that the density be produced by taking an ensemble of density matrices of states $\Psi_i$ of different electron numbers $N_i$, 
\begin{align}  \label{eq:EPPLB}
&E_v[\rho]=\int \rho(\br) v(\br)d\br +F[\rho] \notag \\
&F[\rho] \equiv \min_{\hat{\Dcal} \to \rho} Tr[\hat{\Dcal}(\hat{T}+\hat{W}]  \\
&\hat{\Dcal}= \sum_i \lambda_i \DM{\Psi_i}{\Psi_i}, \, \sum_i \lambda_i=1, \notag \\
&Tr(\hat{\Dcal} \hat{\rho}(\br))=\sum_i \lambda_i\rho_i(\br)=\rho(\br), \notag \\
& \int \rho(\br)d\br=\tilde{N}=\sum_i \lambda_i N_i, \notag
\end{align}
where we indicate general possibly noninteger $N$ with $\tilde{N}$ (see Eschrig \cite{Eschrig1996} for a comprehensive treatment). With the generally assumed convexity of the energy as a function of number of electrons ($E(N) \le \left( E(N+1)+E(N-1))/2\right)$), the minimum energy at $\tilde{N}$ in between, say, the integers $N$ and $N+1$, will be straight line interpolation between $E_0^N$ and $E_0^{N+1}$. So at  $\tilde{N}=N+\omega$ 
\begin{align}\label{eq:N+omega}
&\hat{\Dcal}^{min}=(1-\omega)\DM{\Psi^N_0}{\Psi^N_0} + \omega \DM{\Psi^{N+1}_0}{\Psi^{N+1}_0} \notag \\
&\tilde{E}(\tilde{N})=(1-\omega)E^N_0+\omega E^{N+1}_0  \notag  \\
&\rho(\tilde{N})=(1-\omega)\rho_0^N+\omega \rho^{N+1}_0  
\end{align}
However, $\tilde{E}(\tilde{N})$ cannot be taken to be \textit{the exact energy} of a noninteger electron system at $T=0$. Such a system does not exist, so it is not possible to determine its exact energy. The density $\rho(\tilde{N})$ and energy $\tilde{E}(\tilde{N})$ represent just one out of the many possible ways to continue the density and energy of an $N$-electron system into the nonphysical fractional electron domain. This choice leads to a $\tilde{N}$-derivative with the constant value $-A$ on the $(N,N+1)$ interval, and $-I$ on the $(N-1,N)$ interval. Given the arbitrariness of $\partial E/\partial N$ according to section \ref{sec:Lagrange}, this choice is possible, but it would not work in a density optimization with a Lagrange multiplier for the $\int \rho d\br=N$ constraint, since the derivative is not continuous at $N$. Equations \eqref{eq:EPPLB} and \eqref{eq:N+omega} \textit{for a single $\tilde{N}$-electron system} lack the (statistical) physical interpretation of the straight-line behavior of Refs \cite{PerdewParrLevyBalduz1982,Perdew1985NATO}.  \\
\\
The representation of a fractional system by an  ensemble of density matrices corresponding to different electron numbers leads to a jump in the KS potential when passing the integer $N$, as first noted in \cite{PerdewParrLevyBalduz1982,Perdew1985NATO}. It is known that when we represent the ground state density of a system, $\rho_0^N$ say, with a KS determinant, the exact asymptotic decay of the density according to $e^{-2\sqrt{2I}\,r}$ should be matched by the decay of the KS density, which is that of the slowest decaying orbital density, the HOMO density, $e^{-2\sqrt{-2\epsilon_H}\,r}$. From this follows $\epsilon_H=-I$ \cite{KatrielDavidson1980,AlmbladhBarth1985,LevyPerdewSahni1984}.  If we consider the density $\rho(\tilde{N})$ it is clear that as soon as $\omega>0$ there is a contribution of the ground state $(N+1)$-electron density to $\rho(\tilde{N})$. One can represent the density $\rho(\tilde{N})$ with an ensemble of two KS densities, with the same KS potential and the same orbitals, but with different occupations of the orbitals. One KS density will have  weight $(1-\omega)$ and $N$ occupied orbitals (ignoring spin), and the other one will have weight $\omega$ and $(N+1)$ occupied orbitals. So the $(N+1)$-th orbital (the former LUMO) becomes occupied with $\omega$ electrons, and the slowest decay of the ensemble density will be according to  $e^{-2\sqrt{-2\tilde{\epsilon}_L}\,r}$. Here the orbital energy of the LUMO must be $\tilde{\epsilon}_L=-A$, at any $\omega>0$, since the ionization energy of the negative ion, which dictates the decay of $\rho_0^{N+1}$, is $A$. Now in general the LUMO level of the $N$-electron system, $\epsilon_L$, will be lower than $-A$, therefore the KS potential must shift up by a constant $\Delta$ over the atom or molecule (but not asymptotically) so that the LUMO level is raised to $-A$: $\Delta =-A -\epsilon_L$. If the $N$-electron system was open shell, then HOMO and LUMO would be at the same energy $-I$ (ignoring spin polarization effects), and the upshift $\Delta$ of the potential in this case is $I-A$, equal to the postulated discontinuity $I-A$ in the derivative of the energy at $\tilde{N}=N$, 
\begin{equation}\label{eq:energydiscont}
\left. \frac{\partial \tilde{E}}{\partial \tilde{N}}\right|_{\tilde{N}\downarrow N} - \left. \frac{\partial \tilde{E}}{\partial \tilde{N}}\right|_{\tilde{N}\uparrow N}=I-A=\Delta
\end{equation}
The upshift is therefore also often denoted as ``the derivative discontinuity of the potential", even in the case it is $\Delta =-A -\epsilon_L$.  
If we start with a closed shell system, the LUMO level will be above $\epsilon_H=-I$ and the upshift $\Delta =\tilde{\epsilon}_L-\epsilon_L<I-A$. It is still often denoted as derivative discontinuity. We return to the jump behavior in section \ref{subsec:KSdissociation} \\

 %%%%%%%%%%%%%%%%%%%%%%%%%%%%%%%%%%%%%%%%%%%%%%%%%%%%%%%%%
 \subsection{Locality and size consistency}\label{subsec:Locality}
 %%%%%%%%%%%%%%%%%%%%%%%%%%%%%%%%%%%%%%%%%%%%%%%%%%%%
We have encountered the straight lines (SL) energy as either the average energy according to statistical mechanics of the systems in a macroscopic sample at a properly taken $T \to 0$ limit according to PPLB, or as an arbitrary behavior for a single noninteger electron system at $T=0$ following from a particular construction for the energy in this nonphysical domain, see section \ref{subsec:GrandCanonical}. We call the energy of \eqref{eq:N+omega} the straight-line energy $E^{SL}$. An interesting attempt to prove the linearity of such an energy (without recourse to statistical physics and a grand canonical ensemble) has been made by Yang, Zhang and Ayers (YZA) \cite{YangZhangAyersPRL2000,Ayers2008fractional}. These authors considered the dissociation behavior of
molecules, extending similar arguments by PPLB \cite{PerdewParrLevyBalduz1982} (see next subsection), and find that linearity follows when the functional is required to be local.  Since the property we call ``locality'' is called ``size-consistency'' by YZA, it is important to define terms here.\\ 
Size-consistency is a requirement on any correct theoretical treatment of a set of noninteracting (sub)systems, 
\begin{align}
\text{ Size consistency: }  E(A \cdots B)=E_A+E_B  \label{eq:sizeconsist}  
\end{align}
(the dots indicate long enough distance to make the interaction negligible). 
Size-consistency is a physical requirement formulated for a separation of a system into physical subsystems $A$ and $B$ with defined energies $E_A$ and $E_B$. It does not apply for unphysical fragments lacking a defined energy, such as fragments with a noninteger number of electrons (which are necessarily entangled with other fragments).
Size consistency should be obeyed by any proper functional (any proper quantum chemical method) since the energy is an extensive property.  \\
The property of locality of a functional is something very different. It is not about (physical) systems but about functionals. A DFA is (semi-)local if it computes the energy with an energy density $\epsilon(\br_1)$  that uses at $\br_1$ only $\rho(\br_1)$ or (in the semilocal case) only nearby densities, for instance when derivatives $\nabla\rho(\br_1)$, $\nabla^2\rho(\br_1)$,... are used. LDA, GGA and meta-GGA are examples. We define also a property we call domain-locality. Domain-locality arises when a functional does not determine the energy density at point $\br$ from the electron density at (and in a neighborhood of) $\br$, but determines the energy contributions from separate (nonoverlapping) densities, in cases where the total density is built up from such disjunct pieces. A domain-local functional is for instance the $E^{SL}$ functional for a fractional electron density:  with a density that is separated from the rest of the system, as for instance the atomic densities in dissociated H$_2^+$, it needs integration over that local domain to find the fractional number of electrons $\tilde{N}$, which is an ingredient in the $E^{SL}$ energy determination. For a local functional (either strictly local, semi-local or domain-local)  the following relation holds (the dots indicate nonoverlapping densities)
\begin{align}\label{eq:locality} 
&\text{(strict, semi-, domain-)locality: } \notag \\
 &E^{DFA}[\rho_A \cdots \rho_B]=E^{DFA}[\rho_A]+E^{DFA}[\rho_B]  
\end{align}     
 So locality implies that energy values are assigned even to possibly unphysical (noninteger) densities $\rho_A$ and $\rho_B$.
Nothing in the proof of the HK theorem makes us expect the HK functional to be local. The prime example of nonlocality of the exact functional occurs for a system of two open shell atoms $X$ and $Y$ at large separation in its singlet ground state. The $E_{xc}$ functional cannot be local since its derivative, the xc potential, must be upshifted by a constant over the atom with highest ionization energy, the constant being determined by the ionization energy of the other atom, however remote \cite{AlmbladhvBarth1985,Perdew1985NATO,Gritsenko1996b,GritsenkoBaerends1996,Maitra2005,Maitra2007JCP,HelbigTokatly2009,Maitra2009JCTC,Maitra2012KSpotTDDFT,KraislerKronik2015,Staroverov2016PCCP}. Another example is found when a separated fragment density $\rho_A$ does not correspond to a physical system, for instance when it does not integrate to an integer number of electrons. In that case the exact functional cannot be local. The simplest and very well known case of unphysical fragments is H$_2^+$ at long distance, with electron charge densities of $(1/2)e$ at each site, or an array of $P$ protons with very large distances and 1 electron, with charge densities of $(1/P)e$ at each site. For such a one-electron system the $E_{xc}$ energy has to cancel the Hartree term, i.e.\ $\int \rho(\br) \epsilon_{xc}(\br)d\br = -(1/2)\int \rho(\br_1)\rho(\br_2)/r_{12} d\br_1 d\br_2$, so $\epsilon_{xc}(\br_1)=-(1/2)\int \rho(\br_2)/r_{12} d\br_2$, manifestly nonlocal.  \\
Now one could try to approximate the correct result, for instance for a two-site system, with a (domain-)local functional,
\begin{align}
F[\rho] \overset{?}{=} F^{local}[\rho]=F^{local}[\rho(A)]+F^{local}[\rho(B)]  
\end{align}
This local functional delivers the correct number (but only at infinite distance, outside the range of the Coulomb potential) if it is defined, for fractional electron number,  to be a linear interpolation between the defined functionals for integer electron systems. This is how the $E^{SL}$ functional could be applied. First it is to be considered as a domain-local functional, which is to be applied for each of the two nonoverlapping densities $\rho_A$ and $\rho_B$ separately. These are, in the H$_2^+$ example,  both $1/2$-electron densities. For example, $\rho_A$ is the average of a one-electron density $\rho_H(A)=|1s_H(\br-\mb{R}_A)|^2$ and a zero electron density $\rho_0(A)=0$. For such a local noninteger electron density $E^{SL}$ defines the energy as the average of the one-electron and zero electron energies,
\begin{align}
&F^{SL}[\rho_A \cdots \rho_B] = F^{SL}[\rho(A)]+F^{SL}[\rho(B)] \text{   (locality)}  \notag \\
&F^{SL}[\rho(A)]=F^{SL}\left[\frac{1}{2}\rho_H(A)+\frac{1}{2}\rho_0(A)\right] \notag \\
& \qquad \equiv \frac{1}{2}F[\rho_H(A)]+\frac{1}{2}F[\rho_0(A)]  \label{eq:FlocalAB}
\end{align}
where the first line expresses the requirement that the SL functional is to be applied as a domain-local functional. In the case of $P$ proton sites with one delocalized electron
\begin{align}
F^{SL}[\rho(p)]&=F^{SL}\left[\frac{1}{P}\rho_H(p)+\frac{P-1}{P}\rho_0(p)\right] \notag \\
&  \equiv \frac{1}{P}F[\rho_H(p)]+\frac{P-1}{P}F[\rho_0(p)]  \label{eq:Flocalp}
\end{align}
(with $\rho_H(p)=|1s_H(\br-\mb{R}_p)|^2$ and $\rho_0(p)=0$). Eq.\ \eqref{eq:Flocalp}, where $P$ can be arbitrarily large, shows that the interpolation has to be linear. This does not imply that a fractional electron system is physical. It only shows that we can, in the limit of noninteracting subsystems, try to ignore the nonlocal nature of the exact functional and introduce another functional, not exact, which is taken to be local (as defined in Eq.~\eqref{eq:locality}) and which delivers the correct energy for the total (fully dissociated) system. We have to endow it with the following properties: a) it accepts unphysical fractional electron densities (since such fragment densities may occur); b) it interpolates linearly between the functionals of the nearby integer densities (Eq.~\eqref{eq:Flocalp} with variable $P$ shows that linearity is required \cite{YangZhangAyersPRL2000}). \\
\\
YZA~\cite{YangZhangAyersPRL2000} have demonstrated more generally (for the case of degenerate subsystems)  that the property of locality of a functional (in the case of disjoint subsystems) leads to the requirement that the functional has to be defined for noninteger densities (which the noninteracting subsystems may possess). It is then deduced, as above, that the behavior of the energy for fractional electron densities should be linear. However, we wish to caution that locality is not a property of the exact functional.  The requirement of locality \eqref{eq:locality} is called size consistency in Ref.\ \cite{YangZhangAyersPRL2000}, or functional size consistency. But the two should not be confused: size consistency is the property \eqref{eq:sizeconsist}  (this is called energy size consistency in Ref.\ \cite{YangZhangAyersPRL2000}). Size consistency is a property the exact functional will possess (as any $bona$ $fide$ theory must) and it should be required of approximate functionals (recall the requirement that $A$ and $B$ of Eq.~\eqref{eq:sizeconsist} be physical systems). Locality \eqref{eq:locality} is not a property of the exact functional, and can only be a property of approximate functionals. The requirement of size-consistency, in particular in the case of degenerate subsystems, and the difficulties involved in obeying this requirement  for approximate (local) functionals, have been analyzed by Savin \cite{Savin1996Seminario,Savin2009SizeConsist} and Gori-Giorgi and Savin \cite{GoriG2008SavinJPhysConf}. Locality is not a requirement for size consistency. The exact functional will be nonlocal and must be size consistent.

%%%%%%%%%%%%%%%%%%%%%%%%%%%%%%%%%%%%%%%%%%%%%%%%%%
\subsection{Dissociation into integer electron fragments and (non)locality}\label{subsec:Dissociation}
%%%%%%%%%%%%%%%%%%%%%%%%%%%%%%%%%%%%%%%%%%%%%%%%%
A pleasing property of the SL energy $E^{SL}$ (i.e.\ $\tilde{E}(\tilde{N})$ of \eqref{eq:N+omega}) is that it affords correct dissociation of molecules into integer electron fragments (atoms or molecular fragments), while many LDFA's yield fractionally charged fragments. This problem was originally raised by Slater with the dissociation of NaCl as example, cf.\ Ref.\ \cite{Slater1974Vol4}, Ch.\ 4, and was shown to be solved by a SL energy in Ref.\ \cite{PerdewParrLevyBalduz1982}.  However, it should be recognized that this correct dissociation behavior only follows if the assumption is made that the functional is applicable in local fragments separately (domain-locality). We will review in this section how an $E^{SL}$ leads to dissociation of a diatomic into integer electron atoms. In the next subsection we will then demonstrate that the straight line behavior of the energy still cannot be ``the exact functional''. The derivative discontinuity jump of the KS potential discussed at the end of section \ref{subsec:GrandCanonical} for the case of a single molecule, does not have the right magnitude in the case of dissociation of a heteronucler diatomic. The right jump of the KS potential over an atom in a dissociated system has a nonlocal origin, which is why a locality Ansatz has to fail. \\
Let us denote the energy for fractional electron number $\tilde{E}(\tilde{N})$, where $\tilde{N}=N+\omega, 0 < \omega < 1$ or $-1 < \omega <0$ is the fractional electron number and $\tilde{E}$ is linear on the two intervals. When a system $X-Y$ separates into two parts $X$ and $Y$ with non-overlapping densities $\rho(X)$ and $\rho(Y)$, not necessarily integer,  it is assumed the energy can be obtained as the sum of subsystem energies determined by the subsystem densities: $E[\rho(X \cdots Y)]=\tilde{E}[\rho(X)]+\tilde{E}[\rho(Y)]$ (assumption of locality of the functional).  If  the linear $\tilde{E}$ is applied to each subsystem $X$ and $Y$ separately, dissociation into integer electron systems has the lowest energy. The argument runs as follows. Suppose the starting electron distribution is ``wrong'' in the sense that $A(Y) > I(X)$ (e.g. $Y$=Na$^+$ and $X$=Cl$^-$). With the linearity for the energy at fractional electron number $\tilde{N}$ and constant $\partial \tilde{E}/\partial \tilde{N}=-I$ or $-A$ for $\tilde{N} < N$ and $\tilde{N}>N$ respectively, we will have that a fractional charge transfer $\delta \tilde{N}$ will occur from $X$ to $Y$ with energy change 
\begin{align}\label{eq:DeltaEdissoc}
\Delta E &= \frac{\partial \tilde{E}[X]}{\partial \tilde{N}}(-\delta \tilde{N}) + \frac{\partial \tilde{E}[Y]}{\partial \tilde{N}}(\delta \tilde{N}) \\
&= I(X)\delta \tilde{N} - A(Y)\delta \tilde{N} <0 \notag
\end{align}
If e.g.\ $Y$=Na$^+$, the $3s$ level of Na$^+$ will, due to the derivative discontinuity jump of the KS potential, jump up to the $3s$ level of neutral Na, at $-I$(Na), for an arbitrary amount $\omega$  $(0<\omega < 1)$ of electron transfer, while the $3p_\sigma$ of Cl$^{-1+\omega}$ remains at $-I$(Cl$^-$), see Fig.~\ref{fig:NaCl}a. Since the energy derivatives remain constant at $-A(Y)$ and $-I(X)$ all along the fractional charge on $X$ on the $(N_X,N_X-1)$ interval and on $Y$ on the $(N_Y,N_Y+1)$ interval, transfer will continue until a complete electron has transferred from $X$ to $Y$. So separation of  $X$ and $Y$ will occur into two integer electron systems. However, this is not a stable end result for the SL energy model, see next section. \\
\\
%%%%%%%%%%%%%%%%%%%%%%%%%%%%%%%%%%%%%%%%%%%%%%%%%%%%%
\subsection{Nonlocality and dissociation of an electron pair bond}\label{subsec:KSdissociation}
%%%%%%%%%%%%%%%%%%%%%%%%%%%%%%%%%%%%%%%%%%%%%%%%%%%%%%%

\begin{figure*} %  figure placement: here, top, bottom, or page
\centering
	\includegraphics[width=0.95\columnwidth]{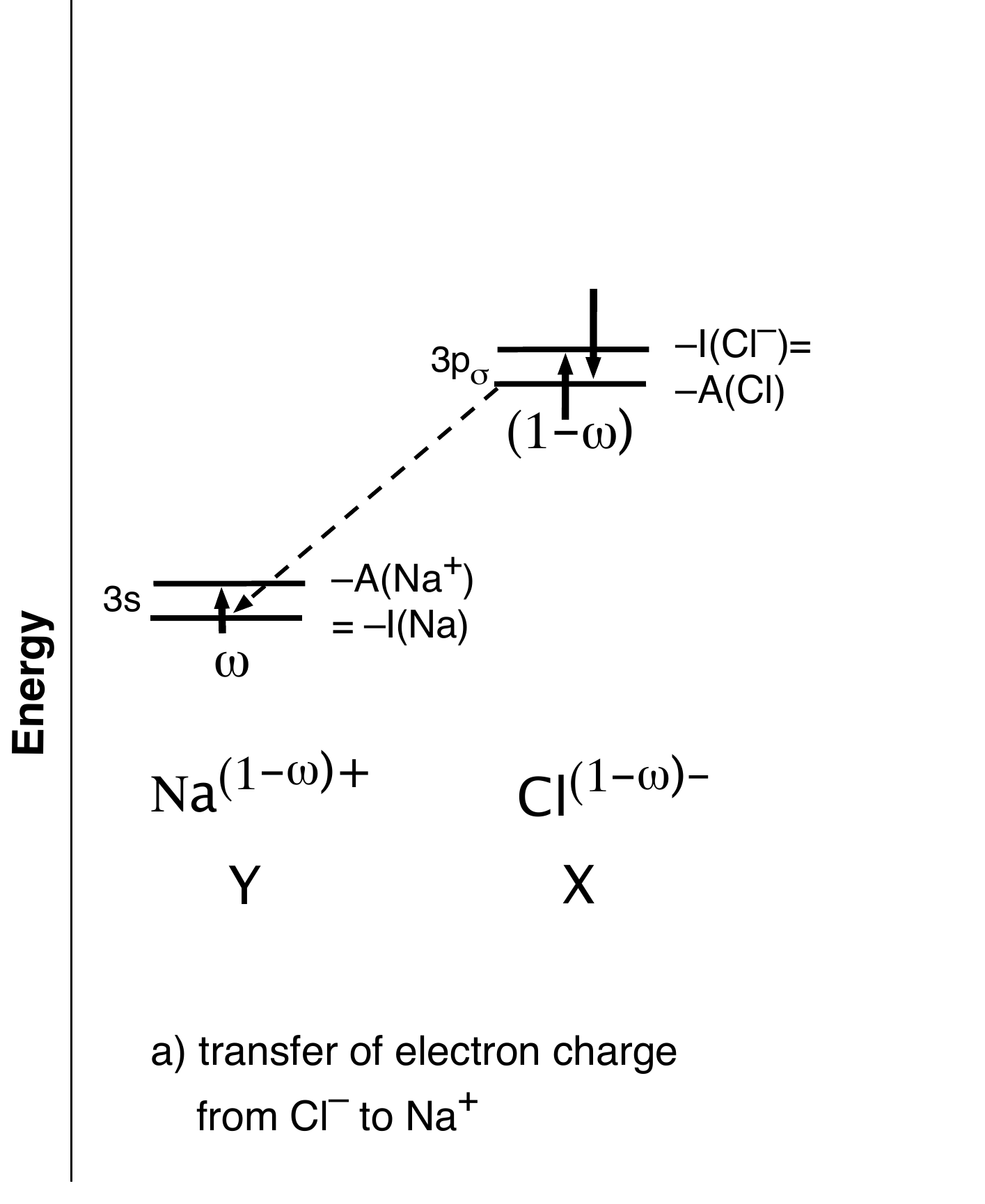}
	\includegraphics[width=1.0\columnwidth]{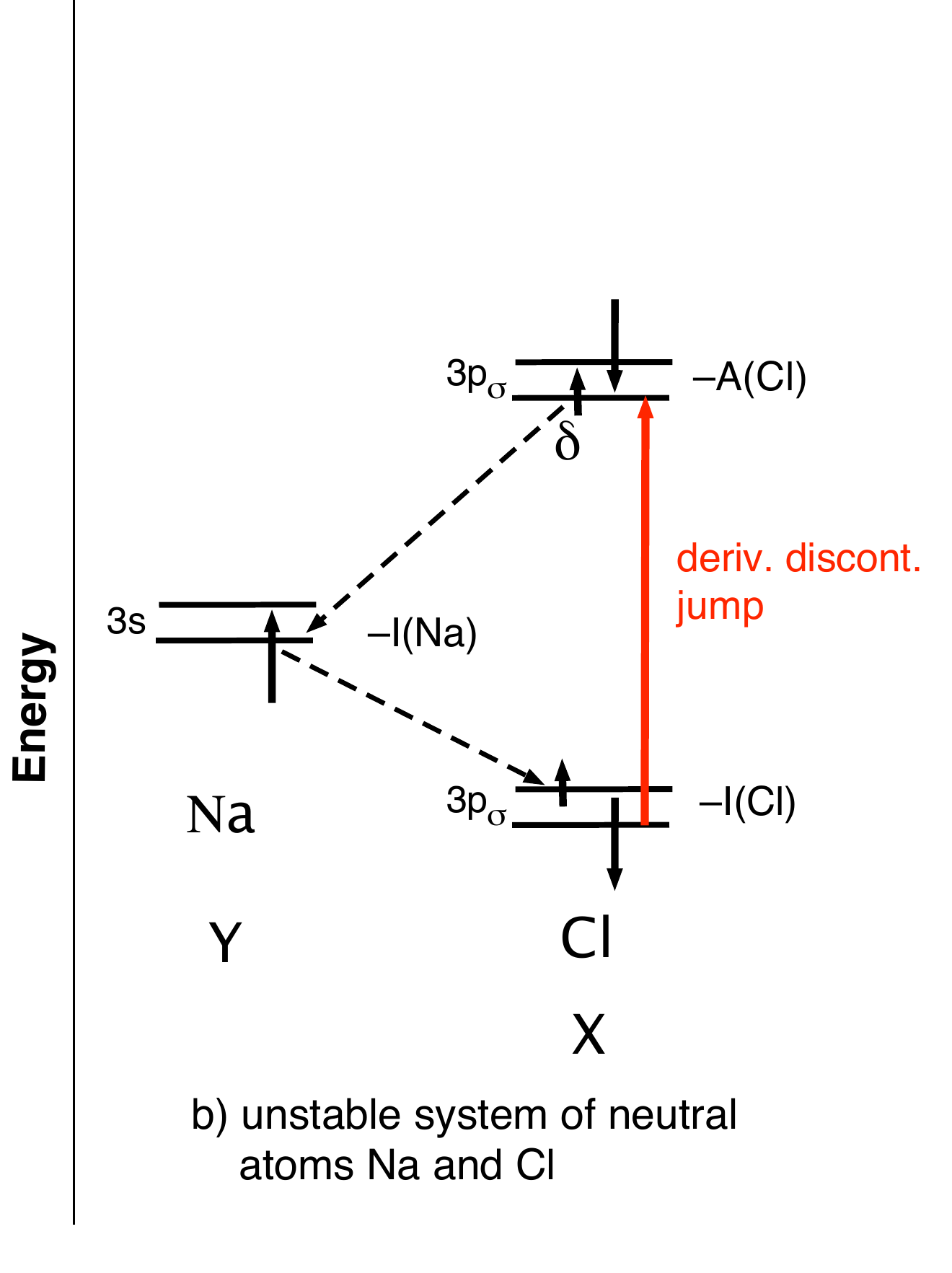}
   \caption{\textbf{(a) Na$^+$Cl$^-$}: The constant frontier orbital energies of Na$^+$ and Cl$^-$ for a straigh-line energy during the charge equalization process from charged fragments to neutral atoms. After the discontinuity jump of the Na$^+$ 3s level has occurred at transfer of any amount $\omega$ $(0<\omega\le1)$ electrons to the 3s level of neutral Na at $-I$(Na), it is still below the $3p_\sigma$ levels of Cl$^-$ ($\alpha$ and $\beta$ level indicated with up and down arrows).\\
\textbf{(b) NaCl}:The dissociated system of two neutral atoms Na and Cl and the derivative discontinuity jump upon a small electron transfer $\delta$ from the Na-3$s\alpha$ orbital to the lower lying Cl-3$p_\sigma\alpha$ orbital. The discontinuity jump puts Cl-3$p_\sigma\alpha$ at $-A$(Cl)=$-I$(Cl$^-$), above the Na-$3s$ level. Degeneracy of Na-3$s\alpha$ and Cl-3$p_\sigma\beta$ does not result.}
\label{fig:NaCl}
\end{figure*}

In spite of the correct dissociation into integer electron fragments, the linear energy behavior and the inherent derivative discontinuity of the KS potential do not lead to completely correct dissociation. The problem is that the linear energy $\tilde{E}$ has to be  applied in combination with a (domain-)locality Ansatz. The systems $X$ and $Y$ are treated as independent subsystems to which locally the linear energy behavior can be applied if they have noninteger electron number.   The typical result of the dissociation discussed so far will be two neutral atoms ($Y$=Na$[(3s\alpha)^1]$, $X$=Cl$[(3p_\pi)^4(3p_\sigma\beta)^1]$ in the given example). But this is not a stable situation. For the neutral atoms $Y$=Na and $X$=Cl we now have that $I(X)>I(Y)$, see Fig.~\ref{fig:NaCl}b. For the atom $X$ the presence of another open shell atom $Y$ far away in the universe, with a smaller ionization potential $I(Y)$ than $I(X)$, makes it necessary that the KS potential exhibits an upshift all over the region of $X$ (but not asymptotically) by the constant $C=I(X)-I(Y)$. This upshift has been identified a long time ago \cite{AlmbladhvBarth1985,Perdew1985NATO} and has received considerable interest \cite{Gritsenko1996b,GritsenkoBaerends1996,Maitra2005,Maitra2007JCP,HelbigTokatly2009,Maitra2009JCTC,Maitra2012KSpotTDDFT,KraislerKronik2015,Staroverov2016PCCP}. Its origin is in the response part of the KS potential \cite{GritsenkoBaerends1996}. That is understandable: the conditional amplitude that is underlying the response potential describes the strong correlation in this case: when one electron is on $Y$ the other electron stays away on $X$ and does not (not even partially) delocalize towards $Y$. The potential step following from the conditional amplitude  arises because of the long range correlation, a manifestly nonlocal effect. It is obvious that this step of the KS potential over atom $X$ requires a nonlocal functional - the presence of another atom $(Y)$ very far away and the magnitude of its ionization energy $I(Y)$ cannot be described with only the local density $\rho(X)$ available.\\
The potential jump $I(X)-I(Y)$ over atom $X$ makes the unpaired spin levels of both atoms degenerate. A closed shell KS molecular orbital solution can result, with doubly occupied HOMO having 50-50 mixture of $X$ and $Y$ character.   This KS solution will have equal probability of spin $\alpha$ or $\beta$ on $Y$ and $X$, in agreement with the exact wavefunction.\\
This correct solution does not result with the standard derivative discontinuity jump of the KS potential. Suppose that the dissociation to neutral atoms described in subsection \ref{subsec:Dissociation} has been completed, see Fig.\ \ref{fig:NaCl}b. As soon as a full electron has been transferred from Cl$^-$, the orbital energies of the generated neutral atoms revert to their ``normal" values ($-I$(Cl) for the $3p_\sigma\beta$ on Cl, $-I$(Na) for $3s\alpha$ on Na). So the empty Cl-$3p_\sigma\alpha$ will be below the occupied Na-$3s\alpha$. Suppose that on a further iteration in an SCF calculation some number $\delta$ of $\alpha$ electrons  is transferred from the higher lying  Na-$3s\alpha$ to the lower lying unoccupied Cl-$3p_\sigma\alpha$ (it could be 1 electron if the transfer is not damped). The ``derivative discontinuity'' jump of the KS potential will put Cl-$3p_\sigma\alpha$ at $-A$(Cl) for $\delta$ electrons. But this will typically be a larger upshift than the required $I$(Cl)-$I$(Na), so this is not a stable situation: the electrons will on the following iteration be sent back to the now lower lying Na-$3s\alpha$ level. If actually carried out, this calculation will lead to infinite oscillation rather than a self-consistent stable solution because the desired situation with 50-50 mixing of $X$(=Cl) and $Y$(=Na) orbitals has to come from perfect degeneracy of $X$ and $Y$ orbitals, which is not achieved if the KS potential plateau of $C=I(X)-I(Y)$ over atom $X$  is not generated.  And it is not generated because the $\tilde{E}(\tilde{N})$ functional lacks the non-locality (i.e.\ knowledge of $Y$) that is required to determine the right upshift of the KS potential over the $X$ atom.  The ``derivative discontinuity'' jump of the KS potential over the $X$ atom of magnitude $I(X)-A(X)$ is too large and prevents a stable proper dissociation situation, see Fig.\ \ref{fig:NaCl}b. The jump should not be $I(X)-A(X)$ (the locally determined derivative discontinuity jump)  but $I(X)-I(Y)$ (dictated nonlocally) in order to obtain proper delocalization of the $\alpha$ and $\beta$ electrons over $X$ and $Y$. \\
\\
The problems discussed in this section and section \ref{subsec:Locality} arise from attempts to treat dissociated systems with neglect of the fact they are still entangled. Entanglement does not show up (or very weakly) in the local densities, but it does in the two-particle density matrix (pair density, conditional amplitude and conditional probabilities for instance), and manifests itself in the nonlocality of the functional.  Nonlocal functionals leading to proper dissociation can be formulated, but they typically require orbital dependency. Refs \cite{Baerends2001,GruningGritsenkoBaerends2003,Chai2012,Chai2014} give examples in a KS context, and in a DMFT context they are commonplace \cite{BuijseBaerends2002,GritsenkoPernalBaerends2005,PirisMatxainLopez2013,LathiotakisHelbig2014PRA,Lathiotakis2015Theophilou,PernalGiesbertz2016}. A genuine density functional treatment of strongly correlated electrons, which is exact in the limiting case of dissociated H$_2$, is provided by the strictly correlated electrons (SCE) functional \cite{Seidl1999,Seidl2007,Gori-Giorgi2015JCTC,GoriGiorgi2016JCTC}. \\

%%%%%%%%%%%%%%%%%%%%%%%%%%%%%%%%%%%%%%%%%%%%%%%%%%%%%%%%%%%%%%%%%%%%%%%%%%%%%
\section{Meaning and applications of the relation $\partial E/\partial n_i=\epsilon_i$ for Hartree-Fock, X$\alpha$ and approximate density functional energies} \label{sec:Slater}
%%%%%%%%%%%%%%%%%%%%%%%%%%%%%%%%%%%%%%%%%%%%%%%%%%%%%%%%%%%%%%%%%%%%%%%%%%

While orbital occupation numbers do not have a place in exact Kohn-Sham DFT, in particular not when noninteger values are considered, this is different in many \textit{approximate} total energy expressions,  corresponding to approximate electronic structure models such as Hartree-Fock, or density functional approximations such as X$\alpha$, LDA, GGAs (LDFAs in general), and hybrid functionals. Even if fractional occupation numbers do not have a physical meaning, they still yield mathematically defined energy values, which allow mathematical manipulations to extract physically meaningful quantities.     
An example is provided by the use of occupation numbers in the Hartree-Fock energy, as originally done by Slater \cite{SlaterMann1969,Slater1972,Slater1974Vol4}, 
\begin{align}\label{eq:EHF1}
&E^{HF1}=\sum_{\substack{\sigma=\alpha,\beta\\p}}  n_p^\sigma \brakket{\phi_p^\sigma}{-\frac{1}{2}\nabla^2+v_{ext}}{\phi_p^\sigma} +\frac{1}{2} \sum_{\substack{\sigma, \tau=\alpha,\beta\\p,q}}  n^\sigma_p n^\tau_q  \\
&\times \int \frac{\phi_p^{\sigma\star}(\bx_1)\phi_q^{\tau\star}(\bx_2)(1-P_{12})\phi_p^{\sigma}(\bx_1)\phi_q^{\tau}(\bx_2)}{|\br_1-\br_2|}  d\bx_1 d\bx_2  \notag 
\end{align}
We use the more general case of unrestricted Hartree-Fock. The summations over $\sigma$ and $\tau$ run over the spin functions $(\alpha,\beta)$. The Hartree-Fock model is defined as the lowest energy determinantal wave function. Orbitals occur in the determinant or not. Fractional occupation numbers have no meaning in the theory, an orbital cannot be partly in the determinant. The introduction of occupation numbers can be seen as a convenient notational device, the occurrence of a spinorbital being determined by the occupation number being 0 or 1. The summations are over all orbitals (we use the conventional notation of indexing occupied orbitals with $i,j,k,l,...$, unoccupied orbitals with $a,b,...$ and general orbitals with $p,q,r,...$). The presence or absence of an orbital is governed by $n_p=1$ or $0$. 
In order to indicate that we extended the Hartree-Fock energy expression with occupation numbers as additional variables, for which we choose linear dependency (power 1), we denote this energy expression as ``HF1". Alternative expressions with occupation numbers are possible, see Eq.~\eqref{eq:EHF2}  below. The energy is usually optimized under orbital variation in $E^{HF1}$ with fixed occupation numbers, where the orthonormality of the orbitals is treated with Lagrange multipliers. Treating the occupation numbers as variables, the optimization has to be carried out under the constraints $\{n_i^\alpha=1, i\le H^\alpha; n_a^\alpha=0, a > H^\alpha \}, \{n_j^\beta=1, j\le H^\beta; n_b^\beta=0, b > H^\beta \}$ where $H^\alpha$ and $H^\beta$ are the highest occupied $\alpha$ and $\beta$ spin orbitals, respectively. It is elementary to derive from Eq.~\eqref{eq:EHF1} the Slater relation
\begin{align}\label{eq:SlaterHF1}
\frac{\partial E^{HF1}}{\partial n_p^\sigma}=\epsilon_p^\sigma
\end{align}
The constraints on the occupation numbers can be treated with Lagrange multipliers because Eq.\eqref{eq:EHF1} defines the objective function $E^{HF1}$ also outside the feasible values of 0 and 1 for the occupation numbers (in contrast to the energy according to HK functional of the density). The Hartree-Fock model itself and the Hartree-Fock energy have no physical meaning outside the feasible values $n_i^\sigma=1, n_a^\sigma=0$. The choice for the value of the objective function $E^{HF}$ outside the feasible domain could be made differently than in $E^{HF1}$, see below. Optimization with Lagrange multipliers for the constraints (including the well-known constraints of orbital orthonormalization) requires stationarity of the Lagrangean
\begin{align}\label{eq:LHF1} 
L^{HF1}&=E^{HF1}-\sum_\sigma \left(\sum_{p,q}\epsilon^\sigma_{pq}\left(\braket{\phi^\sigma_p}{\phi^\sigma_q}-\delta_{pq}\right) \right. \notag \\
&\left. -\sum_{i \le H^\sigma} \lambda^\sigma_i (n^\sigma_i-1)-\sum_{a>H^\sigma} \lambda^\sigma_a (n^\sigma_a-0)  \right) 
\end{align}
The values of the Lagrange multipliers for the occupation number constraints follow from the conditions $\partial L^{HF1}/\partial n^\sigma_p=0$,
\begin{equation}\label{eq:dEHF1dn}
\partial L^{HF1}/\partial n^\sigma_p=0  \quad  \Rightarrow \lambda_p^\sigma=\frac{\partial E^{HF1}}{\partial n_p^\sigma}
\end{equation}
The force of constraint, needed to keep the occupation number at its prescribed value, is proportional to the corresponding Lagrange multiplier $\lambda_p^\sigma$ (in this case equal to it, since $\partial(n_p^\sigma-1)/\partial n_p^\sigma =\partial (n_p^\sigma-0)/\partial n_p^\sigma=1)$. From Eqns \eqref{eq:dEHF1dn} and \eqref{eq:SlaterHF1} follows that
\begin{equation}\label{eq:lambda=epsilon}
\lambda_p^\sigma = \epsilon_p^\sigma
\end{equation}
The fact that the Lagrange multiplier for the constraint of orbital normalization, $\epsilon_p^\sigma$, is equal in magnitude to the force of constraint for constant occupation number, makes physically sense: regardless of the fact that increase of $n_p^\sigma$ to a value $> 1$ does not have physical meaning in the Hartree-Fock model, such increase  in  \eqref{eq:EHF1} would lower $E^{HF1}$, in particular for large negative $\epsilon_p^\sigma$ (core orbitals).  \\
We may recall \cite{Baerends2018JCP} that a different choice could be made for the dependence of the energy on occupation numbers. Since the value of the objective function, $E^{HF}$  outside the domain to which the occupation numbers are constrained is arbitrary, one can choose the function behavior there at will. Of course the force of constraint at the optimum point will depend on the chosen behavior outside the domain, see section \ref{sec:Lagrange}. We can illustrate this by  using for Hartree-Fock  the energy
\begin{align}\label{eq:EHF2}
&E^{HF2}=\sum_{i,\sigma}  (n_i^\sigma)^2 \brakket{\phi_i^\sigma}{-\frac{1}{2}\nabla^2+v_{ext}}{\phi_i^\sigma}  +\frac{1}{2} \sum_{\substack{\sigma, \tau=\alpha,\beta\\i,j}}  (n^\sigma_i n^\tau_j)^2  \\
&\times \int \frac{\phi_i^{\sigma\star}(\bx_1)\phi_j^{\tau\star}(\bx_2)(1-P_{12})\phi_i^{\sigma}(\bx_1)\phi_j^{\tau}(\bx_2)}{|\br_1-\br_2|}  d\bx_1 d\bx_2  \notag 
\end{align}
The total energies $E^{HF1}$ and $E^{HF2}$ are exactly the same at the values $n^\sigma_i=1$ and $n^\sigma_a=0$ (i.e. on the feasible domain), only the behavior outside that domain differs.  Optimization with the same constraint on the occupation numbers being 0 and 1 leads to exactly the same orbitals and orbital energies and the same total energy.  Of course, the derivatives with respect to $n^\sigma_p$ are different, 
\begin{align}
\label{eq:SlaterHF2}
\partial E^{HF2} / \partial n_p^\sigma = 2n_p^\sigma\epsilon_p^\sigma
\end{align}  
Obviously, the force of constraint $\lambda^\sigma_i=2n_i^\sigma\epsilon^\sigma_i$ for occupied orbitals (to keep $n_i^\sigma=1$)  is larger, since the energy now changes with the square of the occupation number (in the unphysical domain outside $n^\sigma_i=1$). The derivative is now larger and then a larger force of constraint is needed to cancel it. The unoccupied orbitals on the other hand now have zero force of constraint since the required zero occupation number is at the minimum of the quadratic $(n_a^\sigma)^2$ behavior for $a \in unocc.$.   \\
\\
It has become popular to move to hybrid functionals with a certain percentage of exact (Hartree-Fock) exchange, in addition to a pure density functional for correlation and the rest of the exchange. When such an energy expression is optimized by orbital variation, as in the traditional Hartree-Fock approach, one will obtain a partly nonlocal potential (a fraction of the Hartree-Fock exchange operator). This procedure is called Hartree-Fock-Kohn-Sham \cite{ParrYang1989} or generalized Kohn-Sham (gKS) \cite{SeidlGorlingVoglMaiewski1996,YangCohenMori2012}. Suppose now that we enter occupation numbers in the exact-exchange part of the energy  in the same way as in Hartree-Fock, and also write the electron density and the kinetic energy and electron-nuclear energies as linear expressions in the occupation numbers, all as in the energy $E^{HF1}$. It is elementary to show that in that case the Slater relation $\partial E^{hybrid1} / \partial n_i = \epsilon_i$ also holds (we use again $hybrid1$ to signal that linear dependence on occupation numbers has been introduced). We recall that the derivation of the Slater relation only uses the chain rule for taking the derivative of the density dependent part $E_{xc}^{DFA}[\rho]$ and, if present, of the one-particle reduced density matrix (1RDM) part $E_{xc}^{DFA}[\gamma]$ of the total energy, cf.\ \cite{Baerends2018JCP}. This relation has also been called the generalized Janak theorem \cite{YangCohenMori2012,PerdewYangBurkeGross2017}. Assume further that a ground state calculation is performed, constraining the occupation numbers to 1 and 0. $[$If we would relieve that constraint and allow fractional occupation numbers, the functional would no longer be distinct from a functional of the one-electron reduced density matrix (1RDM). The exact 1RDM functional has been shown by Gilbert \cite{Gilbert1975} to have complete degeneracy of the orbital energies for all partially occupied orbitals.$]$ If we now consider the derivative with respect to the occupation number of the LUMO, we are dealing with an infinitesimal increase of the total number of electrons. This derivative with respect to occupation number then appears to be the same as the derivative with respect to total electron number, $\partial E^{hybrid1}/ \partial n_L = (\partial E/\partial N)_+$ \cite{MoriCohenYang2008PRL,YangCohenMori2012}. With the straight-line energy the latter is $-A$. This has led to the conclusion that the LUMO orbital energy for an accurate hybrid functional must be close to $-A$. However, this is questionable. In the first place the derivative $\partial E/\partial N$ is not defined in exact DFT, as discussed in section \ref{sec:Lagrange}. It is only $-A$ with a particular choice of the energy behavior for noninteger $N$, namely the straight-line energy $\tilde{E}(\tilde{N})$, Eq.\ \eqref{eq:N+omega}. In the second place, the derivative with respect to occupation number depends on the way occupation numbers are introduced in a DFA. The energy at fractional occupation numbers is a mathematical device, it does not correspond to a true physical system and therefore using it to infer properties of the actual physical system (at integer electron number) is questionable. A LUMO close to $-A$ is typical for the Hartree-Fock model and can for a hybrid be obtained with a very high percentage of exact exchange~\cite{Scheffler2016}. For common hybrids, with exact exchange percentages in the order of 20 - 30\%, the LUMO is much below $-A$ \cite{Baerends2018JCP}. It is remarkable that $in$ $solids$  (but not in molecules) the $\epsilon_L^{DFA}(solid)$ is equal to the DFA calculated (total energy difference) $-A^{DFA}(solid)$ for DFAs that obey the Slater relation \cite{PerdewYangBurkeGross2017}. This difference between solids and molecules is not self-evident. It requires for its proof that the ``solid state limit'' to arbitrarily large size of the system can be taken. This is discussed at some length in Ref.\ \cite{Baerends2018JCP}, see also Ref.\ \cite{PerdewYangBurkeGross2017}.

%%%%%%%%%%%%%%%%%%%%%%%%%%%%%%%%%%%%%%%%%%%%%%%%%%%%%%%%%
\section{Summary}\label{sec:Summary}
%%%%%%%%%%%%%%%%%%%%%%%%%%%%%%%%%%%%%%%%%%%%%%%%%%
We have applied the theory of functional derivatives with constraints and the Lagrange multiplier technique to the problem of optimization of the density in the Hohenberg-Kohn functional with the $\int \rho d\br=N$ constraint. We have highlighted the arbitrariness of a part of the total functional derivative, namely the derivative of the total energy with respect to electron number, $\partial E/\partial N$. It stems from the restricted domain on which $F[\rho]$ is defined. The Hohenberg-Kohn functional $F[\rho]$ (and the Levy-Lieb and Lieb functionals) are only defined on the domain of $N$-electron densities. As a corollary, Janak's theorem does not exist in (exact) Kohn-Sham density functional theory. The essential difficulty is with the nonexistence of fractional electron systems. Any change of an orbital occupation number from integer value (1 or 0) entails a change of electron number from the integer value. There are no wavefunctions for such systems, not for interacting electrons and not for the noninteracting electrons of the Kohn-Sham system. Therefore there are no exact energies and no exchange-correlation energies $E_{xc}$ for noninteger $N$ (at least not for single finite electron systems (atoms and molecules) at $T=0$). \\
For application of the Euler-Lagrange equation to find the optimum $N$-electron density $\rho_0^N$ for which the Hohenberg-Kohn energy functional $E_v[\rho]$ attains its minimum value, the arbitrariness of $\partial E/\partial N$ can be solved in two ways. The functional derivative $\delta E_v[\rho]/\delta \rho(\br)$ should be taken with the constraint $\int \rho(\br) d\br = N$ ($N$ integer). The constraint derivative $\delta E_v[\rho]/\delta_N \rho(\br)$ should be zero at the ground state density $\rho_0^N$. Alternative one can make a suitable choice of the $\partial E/\partial N$ derivative and apply the Lagrange multiplier technique. The choice should be for a continuous and finite derivative, otherwise it would impede the application of the Lagrange multiplier technique for density optimization.  Choices in the literature, such as $F_L[\rho]=+\infty$ in the functional analysis literature of DFT \cite{Lieb1983,Eschrig1996,vanLeeuwen2003,Lammert2007,Kvaal2014} or a discontinuous derivative (different derivatives to the electron rich and electron deficient sides of the integer, are not suitable in this sense. \\
\\ 
It is important that we are dealing here with single quantum systems at $T=0$ (for which the majority of DFT calculations in chemistry are being done).  Perdew et al. \cite{PerdewParrLevyBalduz1982} have studied the statistical mechanics of systems with fluctuating electron number with a constrained search over ensembles of integer electron density matrices, leading to linear energy behavior in between integers for the \textit{average energy} at \textit{average electron number}. It is an altogether different matter to define the energy of a single molecule with a noninteger electron number at $T=0$ by a mixture of density matrices of ground states with different electron numbers, see Eqns \eqref{eq:EPPLB} and \eqref{eq:N+omega}. That procedure represents just one possible extension of the energy functional into the nonphysical fractional electron density domain. Such an extension is essentially arbitrary, cf.\  secion \ref{sec:Lagrange}. The linear choice, which we have denoted $\tilde{E}(\tilde{N})$,  has the interesting property that it leads to the proper dissociation of molecules into neutral atoms, see section \ref{subsec:Dissociation}. We have also emphasized that such success is achieved by treating 
$\tilde{E}(\tilde{N})$ as a local functional, applicable in the local domains of the densities of the fragments separately. This at the same time prevents such a functional from being ``exact'' because fully correct dissociation requires a nonlocal functional, see section \ref{subsec:KSdissociation}. In order to obtain the proper KS wavefunction for a heteronuclear dissociated diatomic molecule $X \cdots Y$ with $I(X)>I(Y)$, the derivative discontinuity of the potential, which is an inherent property of $\tilde{E}(\tilde{N})$, and which would be $I(X)-A(X)$ on atom $X$ (i.e.\ only deopendent on $X$), is wrong; because of nonlocality effects it should be $I(X)-I(Y)$, i.e.\ be dependent on both $X$ and $Y$. \\
\\
Derivatives of the energy with respect to occupation numbers do not exist in exact KS DFT. However, if occupation numbers are introduced as additional variables in the total energy such derivatives can be defined for approximate energy expressions, such as $E^{HF1}$ \eqref{eq:EHF1} and $E^{HF2}$ \eqref{eq:EHF2}, and similarly for X$\alpha$, LDA, GGA and hybrid functionals. At fractional values of these occupation numbers one obtains mathematically defined energies, although without physical meaning. For such energies Slater had already obtained the relation between derivative with respect to orbital occupation number and orbital energy which bears his name and which Janak has tried to extend to (exact) DFT.  The Slater relation is particularly useful when total energies are hard to calculate but orbital energies are readily available \cite{Baerends2018JCP}. These results do not imply that an extension to exact Kohn-Sham DFT is possible. In order to avoid confusion, application of the Slater relation with the approximate DFT energy expressions mentioned above should be denoted as such, and not be called application of Janak's theorem. Janak's theorem is an attempt to extend Slater's relation to exact Kohn-Sham DFT, which has been questioned in section \ref{sec:Janak}.\\
\\
\textbf{Acknowledgement.} I am grateful to Paola Gori-Giorgi and Andreas Savin for helpful discussions and keen interest.

\appendix

\section{N-conserving functional derivatives}\label{sec:derivatives}
A few aspects of constraints on functional derivatives are reviewed. We refer to G\'al \cite{Gal2007,Gal2007JMC} for detailed discussion. In the present case the constraint of conservation of particle number, $\int \rho(\br)d\br=N$ is specifically addressed. 
We follow Parr and Bartolotti \cite{ParrBartolotti1983} in separating the $N$-dependency and shape dependency of the density by writing it as $N$ times a shape function $\sigma(\br)$ which integrates to 1,
\begin{align}\label{eq:rhoNsigma}
&\rho(\br)=N\frac{g(\br)}{\int g(\br')d\br'}=N\sigma(\br); \quad \int \sigma(\br)d\br=1\notag  \\
&\delta\rho(\br)= N\delta\sigma(\br)+\sigma(\br)\delta N = \delta_N\rho(\br)+\delta_\sigma \rho(\br)   \\
&\int \delta_N\rho(\br)d\br =0, \int \delta_\sigma\rho(\br)d\br= \int \sigma(\br)d\br \delta N=\delta N \notag
\end{align}  
where $g(\br)$ is defined up to a scaling constant; it can be $g(\br)=\rho(\br)$. Subscripts $N$ and $\sigma$ denote constant electron number $N$ and constant shape function $\sigma(\br)$, respectively. The constrained derivatives for functionals are analogous to partial derivatives for functions. $\delta \sigma$ should be orthogonal to $\sigma$ otherwise the $N\delta\sigma$ term contains part of the electron number variation, which should follow from $\sigma \delta N$ (if $\delta\sigma$ has a component $a\sigma$ along $\sigma$, $\int N\delta\sigma d\br = \int Na\sigma d\br =Na$).  \\
Define the  changes $\delta_N E[\rho]$ of the functional $E[\rho]$ as those variations of the energy that are induced by $N$-conserving density changes $\delta_N\rho(\br)$, $\int (\rho(\br)+\delta_N\rho(\br))d\br=N$,
\begin{equation}\label{eq:defdeltaNE}
\delta_N E[\rho] \equiv \int \frac{\delta E}{\delta\rho(\br)} \delta_N\rho(\br)d\br.
\end{equation}
Application of the Lagrange multiplier technique requires variation of $E$ to be defined for arbitrary variations $\delta \rho$, so if we want to break down the total derivative $\delta E/\delta \rho$ into norm-conserving and shape-conserving parts, these have to be defined as operators acting on arbitrary $\delta \rho$,
\begin{align}
\int \frac{\delta E}{\delta \rho}\delta\rho d\br =\int  \left(  \frac{\delta E}{\delta_N\rho(\br)} +\frac{\delta E}{\delta_{\sigma}\rho(\br)} \right) \delta\rho(\br)d\br
\end{align}  
The constraint derivative $\delta E[\rho]/\delta_N\rho(\br)$ is defined as that operator (kernel of an integral operator) that delivers $\delta_N E$ from a general $\delta\rho$, by picking out the $\delta_N\rho(\br)$ part of $\delta\rho(\br)$ 
\begin{align}\label{eq:delNE}
\delta_N E&=\int \frac{\delta E}{\delta_N\rho(\br)} \delta\rho(\br)d\br=\frac{\delta E}{\delta_N\rho(\br)} (\delta_N\rho(\br)+\delta_\sigma\rho(\br))d\br \notag \\
&=\int  \frac{\delta E}{\delta_N\rho(\br)} \delta_N\rho(\br)d\br + \int \frac{\delta E}{\delta_N\rho(\br)} \delta_\sigma\rho(\br)d\br \notag \\
&= \int  \frac{\delta E}{\delta_N\rho(\br)} \delta_N\rho(\br)d\br + 0.
\end{align}
The fact that the last term on the second line is zero is an essential element in the definition of $\delta E/\delta_N\rho$ - the total $\delta_N E$ stems from and is determined only by the $N$-conserving part of $\delta\rho$. The shape-conserving part, which only allows a change of  $N$, cannot contribute to $\delta_N E$. This is an important point. In the literature \cite{ParrBartolotti1983,PerdewLevy1983,ParrYang1989,ParrLiu1997} it has been an issue that $\delta E[\rho]/\delta_N \rho(\br)$ would be ambiguous, containing an undefined constant. One must then be careful to note that this stems from restricting the domain on which the derivative is defined to the set of $N$-conserving density changes $\delta_N\rho(\br)$ and the observation that $\int C\delta_N\rho(\br) d\br=0$ \cite{Gal2001,vanLeeuwen2003}.  However, in the application of the Lagrange multiplier method (see text) we need derivatives (operators) that are defined for action on an arbitrary $\delta \rho$, not just on the domain of $N$-conserving $\delta_N\rho$. So ``orthogonality" of $\delta E/\delta_N\rho(\br)$ on $\delta_\sigma\rho(\br)$, $\int (\delta E/\delta_N\rho(\br))\delta_\sigma\rho(\br)d\br=0$, is a necessary part of the definition of $\delta E/\delta_N\rho(\br)$. This property would be destroyed by an additional constant $C$: $\int C\delta_\sigma\rho(\br)d\br=C\delta N$. A constant would give $\delta E/\delta_N\rho(\br)$ an unwarranted component in the space ``orthogonal" to the set $\{\delta_N\rho(\br)\}$. This is analogous to adding to $\mb{\nabla}_{//}f$ in Fig.~\ref{fig:LagrangeMultiplier} a component $C\mb{n}_\bot$ in the direction $\mb{n}_\bot$ perpendicular to $\mb{\nabla}_{//}f$, so it would clutter $df=(\mb{\nabla}_{//}f+C\mb{n}_\bot)\centerdot \delta\mb{l}$ for a general displacement $\delta\mb{l}$ with contributions $C\mb{n}_\bot \centerdot \delta \mb{l}_{\bot}$ from the perpendicular component $\delta \mb{l}_{\bot}$.  \\
In the same way, for a complete definition of $\delta E/\delta_{\sigma}\rho$  we must have $\int (\delta E/\delta_\sigma\rho(\br))\delta_N\rho(\br)d\br=0$. Now for arbitrary $\delta\rho$ 
\begin{align}
&\int \frac{\delta E}{\delta\rho(\br)}\delta\rho(\br)d\br  \notag \\
&=\int \frac{\delta E}{\delta\rho(\br)}\delta_N\rho(\br)d\br + \int \frac{\delta E}{\delta\rho(\br)}\delta_\sigma \rho(\br)d\br \notag \\
&=\int \frac{\delta E}{\delta_N\rho(\br)}\delta_N\rho(\br)d\br + \int \frac{\delta E}{\delta_{\sigma}\rho(\br)}\delta_\sigma \rho(\br)d\br \notag \\
&=\int \frac{\delta E}{\delta_N\rho(\br)}\delta\rho(\br)d\br + \int \frac{\delta E}{\delta_{\sigma}\rho(\br)}\delta\rho(\br)d\br    \\
&=\int  \left(\frac{\delta E}{\delta_N\rho(\br)} + \frac{\delta E}{\delta_\sigma\rho(\br)}\right) \delta\rho(\br)d\br, \notag 
\end{align}
consistent with the identity
\begin{align}\label{eq:dEdN+dEsigma}
&\frac{\delta E}{\delta\rho(\br)}=\frac{\delta E}{\delta_N\rho(\br)}+\frac{\delta E}{\delta_\sigma\rho(\br)}.
\end{align}
Now a derivative with constant shape function $\sigma$ can only involve a change of $N$, so we expect
\begin{align}
\frac{\delta E}{\delta_\sigma\rho(\br)}=\frac{\partial E}{\partial N} 
\end{align}
and with the chain rule
\begin{align}\label{eq:partialEpartialN}
\frac{\partial E}{\partial N}=\int \frac{\delta E}{\delta \rho(\br)} \frac{\partial \rho(\br)}{\partial N} d\br = \int \frac{\delta E}{\delta \rho(\br)} \frac{\rho(\br)}{N}d\br.
\end{align}
A derivation can be given  for $\delta E[\rho]/\delta_N\rho(\br)$ employing the factorisation of $\rho(\br)$ as $Ng(\br)/\int g(\br')d\br'$ (see \cite{Gal2001}) which leads to 
\begin{align}\label{eq:dEdNrho}
\frac{\delta E[\rho]}{\delta_N\rho(\br)}=\frac{\delta E[\rho]}{\delta\rho(\br)}- \int \frac{\delta E[\rho]}{\delta \rho(\br)} \frac{\rho(\br)}{N}d\br,
\end{align}
in perfect agreement with Eq. \eqref{eq:partialEpartialN}, so that with Eqns \eqref{eq:dEdN+dEsigma} - \eqref{eq:dEdNrho}
\begin{equation}
\frac{\delta E[\rho]}{\delta\rho(\br)} = \frac{\delta E[\rho]}{\delta_N\rho(\br)} + \frac{\partial E}{\partial N}. 
\end{equation}
Note that indeed $\int (\delta E/\delta_\sigma\rho(\br))\delta_N\rho(\br)d\br=\int (\partial E/\partial N)\delta_N\rho(\br)d\br = (\partial E/\partial N)N\int \delta\sigma(\br)d\br=0$. The fact that $\partial E/\partial N$ is not defined by the Hohenberg-Kohn theorem, is not a problem in DFT, since only the energy changes $\delta_N E[\rho]$ are required. The fact that any constant $C$ has no effect when acting on a feasible $\delta_N\rho(\br)$ makes sure that the essential change $\delta_N E[\rho]$ is defined whatever the (chosen) value of $\partial E/\partial N$:
\begin{align}
&\int \frac{\delta E[\rho]}{\delta\rho(\br)} \delta_N\rho(\br) = \int \left(\frac{\delta E[\rho]}{\delta_N\rho(\br)} + \frac{\partial E}{\partial N}\right)\delta_N\rho(\br) \notag  \\
&=  \int  \frac{\delta E[\rho]}{\delta_N\rho(\br)}\delta_N\rho(\br)=\delta_N E[\rho].
\end{align}\\
\\

\section{An arbitrary constant in the constrained derivative?}\label{sec:ParrYang}
Starting with the contribution by Parr and Bartolotti \cite{ParrBartolotti1983} remarks have been made in the literature that the constrained derivative
\begin{equation}
\left[ \frac{\delta E[\rho]}{\delta \rho(\br)}\right]_N
\end{equation} 
which we have written for the sake of concise notation as $\delta E[\rho] / \delta_N \rho(\br)$, following G\'al \cite{Gal2001},  would only be defined up to an arbitrary constant.  The origin is the same pitfall we mentioned in section \ref{sec:Lagrange}. Parr and Yang (\cite{ParrYang1989}, p. 83) write Eq.\ \eqref{eq:dEdg1} with an $N$-conserving derivative 
\begin{equation}\label{eq:dEdgN}
\frac{\delta E[\rho]}{\delta g(\br)} =\int \left[ \frac{\delta E[\rho]}{\delta \rho(\br')}\right]_N \frac{\delta \rho(\br')}{\delta g(\br)}d\br'=0
\end{equation}
where $[\delta E[\rho]/\delta \rho(\br)]_N$ is ``a functional differentiation with $N$ fixed" (what we denote $\delta E/\delta_N\rho(\br)$). Application of the chain rule means that Eq.~\eqref{eq:dEdgN} should in principle be written with the full derivative $\delta  E[\rho_0^N]/\delta \rho(\br')$, as has been done in Eq.~\eqref{eq:dEdg1}, but since $\delta \rho(\br')/\delta g(\br)$ will deliver $N$-conserving variations $\delta_N\rho(\br')$, the use of the constraint derivative in this case is also correct. Using Eqns \eqref{eq:drhodg} and \eqref{eq:dEdgN} at the solution point $(\rho_0^N)$ one then obtains  equation (4.4.7) of \cite{ParrYang1989}, p. 83,
\begin{equation}\label{eq:dErho0drhoN}
\left[ \frac{\delta E[\rho_0^N]}{\delta \rho(\br)}\right]_N =\int  \left[\frac{\delta E[\rho_0^N]}{\delta \rho(\br')}\right]_N \frac{g(\br')}{\int g(\br'')d\br''}d\br'
\end{equation}
from which it was concluded that $[\delta E[\rho_0^N]/\delta \rho(\br)]_N$ must be an undefined constant (the constant cannot be fixed from \eqref{eq:dErho0drhoN}). However, we have seen in section \ref{sec:Lagrange} that at the solution point this $N$-conserving derivative is not an arbitrary constant (the pitfall mentioned there) but is zero. There is not a problem with Eq.~\eqref{eq:dErho0drhoN}, it is correct, but one should be careful with conclusions drawn from it.  It is possible to conclude that the $N$-conserving derivative at the solution point is a constant, but it may be that on other grounds it is known that this constant is not arbitrary but has a specific value. The constrained derivative has been found to be zero at the solution point $\rho_0^N$, see section \ref{sec:Lagrange}. That is of course perfectly consistent with \eqref{eq:dErho0drhoN}. If on the other hand the full derivative would have been used in Eq.~\eqref{eq:dEdgN}, as we did in Eq.~\eqref{eq:dEdg1}, this would also have emerged in Eq.~\eqref{eq:dErho0drhoN}. We have noted in section \ref{sec:Lagrange} that it is this full derivative that is equal to a constant at $\rho_0^N$, which is again compatible with equation  \eqref{eq:dErho0drhoN} (for the full derivative $\delta E[\rho_0^N]/\delta \rho(\br)$ instead of the constrained $\delta E[\rho_0^N]/\delta_N \rho(\br)$). As we have seen, the constant in that case is indeed arbitrary, being $\partial E[\rho_0^N]/\partial N$. This difference - the constrained derivative $\delta E[\rho_0^N] /\delta_N\rho(\br)$ being zero according to our present analysis, see discussion below Eq.\ \eqref{eq:dErho0drho}, but equal to an undefined constant according to Ref.~\cite{ParrYang1989} - goes back to Parr and Bartolotti \cite{ParrBartolotti1983}, who first proposed there to be an arbitrary constant in $\left[ \delta E[\rho_0^N]/\delta\rho(\br) \right]_N$, which has since often surfaced in the DFT literature.\\

%\bibliography{../../../Libraries/biblioTCVU2016}

\end{document}